\documentclass[a4paper,fleqn,usenatbib]{mnras}

\usepackage[T1]{fontenc}
\usepackage{ae,aecompl}

\usepackage{graphicx}	
\usepackage{amsmath}	
\usepackage{amssymb}	

\renewcommand{\theenumi}{\arabic{enumi})}

\title[Magnetic fields in eccentric TDE Discs]{Importance of magnetic fields in highly eccentric discs with applications to tidal disruption events}

\author[E. M. Lynch and G. I. Ogilvie]{
Elliot M. Lynch \thanks{E-mail: eml52@cam.ac.uk} and Gordon I. Ogilvie
\\
Department of Applied Mathematics and Theoretical Physics, University of Cambridge, Centre for Mathematical Sciences, \\ Wilberforce Road, Cambridge CB3 0WA, UK\\
}

\date{Accepted XXX. Received YYY; in original form ZZZ}

\pubyear{2019}

\begin{document}
\label{firstpage}
\pagerange{\pageref{firstpage}--\pageref{lastpage}}
\maketitle

\begin{abstract}
Whether tidal disruption events (TDEs) circularise or accrete directly as a highly eccentric disc is the subject of current research and appears to depend sensitively on the disc thermodynamics. In a previous paper we applied the theory of eccentric discs to TDE discs using an $\alpha-$prescription for the disc stress, which leads to solutions that exhibit extreme, potentially unphysical, behaviour. In this paper we further explore the dynamical vertical structure of highly eccentric discs using alternative stress models that are better motivated by the behaviour of magnetic fields in eccentric discs. We find that the presence of a coherent magnetic field has a stabilising effect on the dynamics and can significantly alter the behaviour of highly eccentric radiation dominated discs. We conclude that magnetic fields are important for the evolution of TDE discs.

\end{abstract}




\section{Introduction}

Tidal disruption events (TDEs) are transient phenomena where an object on a nearly parabolic orbit passes within the tidal radius and is disrupted by the tidal forces, typically a star being disrupted by a supermassive black hole (SMBH). Bound material from the disruption forms a highly eccentric disc, which in the classic TDE model of \citet{Rees88} are rapidly circularised as the material returns to pericentre. It has, however, been proposed that circularisation in TDEs may be inefficient resulting in the disc remaining highly eccentric \citep{Guillochon14,Piran15,Krolik16,Svirski17}. In \citet{Lynch20} (henceforth Paper I) we presented a hydrodynamical model of these highly eccentric discs applied to TDEs where circularisation is inefficient.

Two issues were highlighted in Paper I. One was confirming that radiation pressure dominated discs are thermally unstable when the viscous stress scales with total pressure in highly eccentric discs, a result that has long been known for circular discs \citep{Shakura76,Pringle76,Piran78}. Circular radiation pressure dominated discs can be stabilised by assuming stress scales with gas pressure \citep{Meier79,Sakimoto81}.  For highly eccentric discs it appears that the thermal instability is still present when stress scales with gas pressure; however there exists a stable radiation pressure dominated branch which is the outcome of the thermal instability. For typical TDE parameters, this branch is very hot and often violates the thin-disc assumptions.

The second issue was the extreme behaviour that can occur during pericentre passage. For the radiation pressure dominated disc where stress scales with gas pressure the solution is nearly adiabatic and undergoes extreme compression near pericentre. In models where the viscous stresses contribute to the dynamics we typically found that the vertical viscous stress is comparable to or exceeds the (total) pressure, which is possibly problematic for the $\alpha$-model as it would indicate transonic turbulence. In some of the solutions the vertical viscous stress can exceed the total pressure by an order of magnitude, strongly violating the assumptions of the $\alpha$-model. 

In this paper we focus on the second of the two issues by considering alternative turbulent stress models which are better motivated by the physics of the underlying magnetic field to see if this resolves some of the extreme behaviour seen in the $\alpha$-models. We will also see if alternative stress models are more thermally stable than the $\alpha$-model, although it's possible the solution to this issue is outside the scope of a thin disc model .

Two additional physical effects, not present in an $\alpha$-model, may be important for regulating the extreme behaviour at pericentre. One is the finite response time of the magnetorotational instability (MRI) (see for instance the viscoelastic model of \citet{Ogilvie01} and discussion therein) which means the viscous stress cannot respond instantly to the rapid increase in pressure and velocity gradients during pericentre passage, potentially weakening the viscous stresses so they no longer exceed the pressure. Another is the relative incompressibility of the magnetic field, compared with the radiation or the gas, with the magnetic pressure providing additional support during pericentre passage which could prevent the extreme compression seen in some models.

Various attempts have been made to rectify some of the deficiencies of the $\alpha-$prescription using alternative closure models for the turbulent stress. \citet{Ogilvie00,Ogilvie01} proposed a viscoelastic model for the dyadic part of the Maxwell stress (i.e.\ the contribution from magnetic tension $\frac{B^{i} B^{j}}{\mu_0}$) to account for the finite response time of the MRI. It was shown in \citet{Ogilvie03b} that for incompressible fluids there is an exact asymptotic correspondence between MHD in the limit of large magnetic Reynolds number and viscoelastic fluids (specifically an Oldroyd-B fluid) in the limit of large relaxation time. \citet{Ogilvie02} improved upon the compressible viscoelastic model of \citet{Ogilvie00,Ogilvie01} by including an isotropic part to the stress to model the effects of magnetic pressure and correcting the heating rate so that total energy is conserved. \citet{Ogilvie03} proposed solving for both the Maxwell and Reynolds stresses and suggested a nonlinear closure model based on requiring the turbulent stresses to exhibit certain properties (such as positive definiteness, and relaxation towards equipartition and isotropy) known from simulations and experiments. 

Simulations of MRI in circular discs typically find that the magnetic pressure tends to saturate at about 10\% of the gas pressure. However in the local linear stability analysis of \citet{Pessah05} the toroidal magnetic field only stabilises the MRI when it is highly suprathermal (specifically when the Alfv\'{e}n speed is greater than the geometric mean of the sound speed and the Keplerian speed). \citet{Das18} confirmed this result persists in a global linear eigenmode calculation. In light of this, \citet{Begelman07} have suggested that, for a strongly magnetised disc, the viscous stress may scale with the magnetic pressure and showed that such a disc would be thermally stable even when radiation pressure dominates over gas pressure. Such a disc was simulated by \citet{Sadowski16}, who indeed found thermal stability.


Throughout this paper we will make use of certain conventions from tensor calculus, such as the Einstein summation convention and the distinction between covariant and contravariant indices, along with the notation for symmetrising indices,

\begin{equation}
X^{(i j)} := \frac{1}{2} \left(X^{ i j} + X^{j i} \right) .
\end{equation}

This paper is structured as follows. In Section \ref{orbital coords} we discuss the geometry of eccentric discs and restate the coordinate system of \citet{Ogilvie19}. In Section \ref{breathinmode} we derive the equations for the dynamical vertical structure, including the effects of a Maxwell stress, in this coordinate system. In Section \ref{magnetic pressure effect} we consider a model with an $\alpha-$viscosity and a coherent magnetic field which obeys the ideal induction equation. In Section \ref{constituative model} we consider a nonlinear constitutive model for the magnetic field. In our discussion we discuss the stability of our solutions (Section \ref{stabil}) and the possibility of dynamo action in the disc  (\ref{dynamo}). We present our conclusions in Section \ref{conc} and additional mathematical details are in the appendices.

\section{Orbital Coordinates} \label{orbital coords}

Similar to Paper I we assume the dominant motion in a TDE disc consists of elliptical Keplerian orbits, subject to relatively weak perturbations from relativistic precessional effects, pressure and Maxwell stresses. This model is unlikely to be applicable to TDEs where the tidal radius is comparable to the gravitational radius owing to the strong relativistic precession.

Let $(r,\phi)$ be polar coordinates in the disc plane. The polar equation for an elliptical Keplerian orbit of semimajor axis $a$, eccentricity $e$ and longitude of periapsis $\varpi$ is

\begin{equation}
r = \frac{a (1 - e^2)}{1 + e \cos f} \quad ,
\end{equation} 
where $f = \phi - \varpi$ is the true anomaly. A planar eccentric disc involves a continuous set of nested elliptical orbits. The shape of the disc can be described by considering $e$ and $\varpi$ to be functions of $a$. The derivatives of these functions are written as $e_a$ and $\varpi_a$, which can be thought of as the eccentricity gradient and the twist, respectively. The disc evolution is then described by the slow variation in time of the orbital elements $e$ and $\varpi$ due to secular forces such as pressure gradients in the disc and departures from the gravitational field of a Newtonian point mass.

In this work we adopt the (semimajor axis $a$, eccentric anomaly $E$) orbital coordinate system described in \citet{Ogilvie19}. The eccentric anomaly is related to the true anomaly by

\begin{equation}
\cos f = \frac{\cos E - e}{1 - e \cos E} , \quad  \sin f = \frac{\sqrt{1 - e^2} \sin E}{1 - e \cos E}
\end{equation}
and the radius can be written as

\begin{equation}
r = a (1 - e \cos E) \quad .
\end{equation}

The area element in the orbital plane is given by $d A = (a n/2) J \, d a \, d E$ where $J$ is given by

\begin{equation}
J = \frac{2}{n} \left[\frac{1 -e (e + a e_a)}{\sqrt{1 - e^2}} - \frac{a e_a}{\sqrt{1 - e^2}} \cos E - a e \varpi_a \sin E\right] ,
\end{equation} 
which  corresponds to the Jacobian of the $(\Lambda,\lambda)$ coordinate system of \citet{Ogilvie19}. Here $n = \sqrt{\frac{G M_{\bullet}}{a^3}}$ is the mean motion with $M_{\bullet}$ the mass of the black hole. The Jacobian can be written in terms of the orbital intersection parameter $q$ of \citet{Ogilvie19}:

\begin{equation}
J = (2/n) \frac{1 -e (e + a e_a)}{\sqrt{1 - e^2}}  (1 - q \cos(E - E_0))
\end{equation}
where $q$ is given by

\begin{equation}
q^2 = \frac{(a e_a)^2 + (1 - e^2) (a e \varpi_a)^2}{[1 - e (e + a e_a)]^2} \quad ,
\end{equation}
and we require $|q| < 1$ to avoid an orbital intersection \citep{Ogilvie19}. The angle $E_0$, which determines the location of maximum horizontal compression around the orbit, is determined by the relative contribution of the eccentricity gradient and twist to $q$:

\begin{equation}
\frac{a e_a}{1 - e(e + a e_a)} = q \cos E_0 \quad. 
\end{equation}

Additionally it can be useful to rewrite time derivatives, following the orbital motion, in terms of the eccentric anomaly:

\begin{equation}
\frac{\partial}{\partial t} = \frac{n}{(1 - e \cos E)} \frac{\partial}{\partial E} \quad .
\end{equation}

\section{Derivation of the equations of vertical structure, including thermal effects} \label{breathinmode}

A local model of a thin, Keplerian eccentric disc was developed in \citet{Ogilvie14}. In Paper I we developed a purely hydrodynamic model, which included an $\alpha$-viscosity prescription along with radiative cooling, allowing for contributions to the pressure from both radiation and the gas, in the $(a,E)$ coordinate system of \citet{Ogilvie19}. In a similar vein we here develop a local model that allows for a more general treatment of the turbulent/magnetic stress.

The equations, formulated in a frame of reference that follows the elliptical orbital motion, are the vertical equation of motion,

\begin{equation}
\frac{D v_z}{D t} = - \frac{G M_{\bullet} z}{r^3} - \frac{1}{\rho} \frac{\partial}{\partial z} \left(p + \frac{1}{2} M - M_{z z} \right),
\label{vertical laminar flow}
\end{equation}
the continuity equation,

\begin{equation}
\frac{D \rho}{D t} = - \rho \left( \Delta + \frac{\partial v_z}{\partial z} \right) ,
\end{equation}
and the thermal energy equation, 

\begin{equation}
\frac{D p}{D t} =  - \Gamma_1 p \left( \Delta + \frac{\partial v_z}{\partial z} \right) + (\Gamma_3 - 1) \left( \mathcal{H} - \frac{\partial F}{\partial z} \right) ,
\end{equation}
where, for horizontally invariant ``laminar flows'',

\begin{equation}
\frac{D}{D t} = \frac{\partial}{\partial t} + v_z \frac{\partial}{\partial z}
\end{equation}
is the Lagrangian time-derivative,

\begin{equation}
\Delta = \frac{1}{J} \frac{dJ}{dt}
\end{equation}
is the divergence of the orbital velocity field, which is a known function of $E$ that depends of $e$, $q$ and $E_0$. $F = F_{\rm rad} + F_{\rm ext}$ is the total vertical heat flux with

\begin{equation}
F_{\rm rad} = - \frac{16 \sigma T^3}{3 \kappa \rho} \frac{\partial T}{\partial z}
\end{equation}
being the vertical radiative heat flux and $F_{\rm ext}$ containing any additional contributions to the heat flux (such as from convection or turbulent heat transport). The tensor

\begin{equation}
M^{i j} := \frac{B^i B^j}{\mu_0}
\end{equation}
is the part of the Maxwell stress tensor arising from magnetic tension. This can include contributions from a large scale mean field and from the disc turbulence. Its trace is denoted $M = M^{i}_{\,\,\, i}$, which corresponds to twice the magnetic pressure. In this paper we shall explore two different closure models for the time-evolution of $M^{ij}$.

Following Paper I, we write the heating rate per unit volume, resulting from the dissipation of magnetic/turbulent energy, as

\begin{equation}
\mathcal{H} = f_{\mathcal{H}} n p_v ,
\end{equation}
where $f_{\mathcal{H}}$ is a dimensionless expression that depends on the closure model and $p_v$ is a pressure to be specified in the Maxwell stress closure model.

In addition to the magnetic pressure, which is included through the $\frac{1}{2} M$ term in equation (10), the pressure includes contributions from radiation and a perfect gas with a ratio of specific heats $\gamma$. We define the hydrodynamic pressure to be the sum of the gas and radiation pressure,

\begin{equation}
p = p_{r} + p_{g} = \frac{4 \sigma}{3 c} T^4 + \frac{\mathcal{R} \rho T}{\mu},
\end{equation}
and $\beta_r$ to be the ratio of radiation to gas pressure:

\begin{equation}
\beta_{r} := \frac{p_r}{p_g} = \frac{4 \sigma \mu}{3 c \mathcal{R}} \frac{T^3}{\rho} \quad .
\end{equation}
We assume a constant opacity law, applicable to the electron-scattering opacity expected in a TDE, with the opacity denoted by $\kappa$.

We consider a radiation+gas mixture where $F_{\rm ext}$ is assumed to be from convective or turbulent mixing and the first and third adiabatic exponents are given by \citep{Chandrasekhar67}

\begin{equation}
\Gamma_1 = \frac{1 + 12 (\gamma - 1) \beta_r + (1 + 4 \beta_r)^2 (\gamma - 1)}{(1 + \beta_r) (1 + 12 (\gamma - 1) \beta_r)} \quad ,
\label{Gamma 1 radmix}
\end{equation}

\begin{equation}
\Gamma_3 = 1 + \frac{(1 + 4 \beta_r) (\gamma - 1)}{1 + 12 (\gamma - 1) \beta_r} \quad .
\end{equation}

As in Paper I, we propose a separable solution of the form

\begin{align}
\begin{split}
\rho &= \hat{\rho} (t) \tilde{\rho} (\tilde{z}) , \\
p &= \hat{p} (t) \tilde{p} (\tilde{z}) , \\
M &= \hat{M}_{i j} (t) \tilde{M} (\tilde{z}) , \\
F &= \hat{F} (t) \tilde{F} (\tilde{z}) , \\
v_{z} &= \frac{d H}{d t}\tilde{z} , \\
\end{split}
\end{align}
where 

\begin{equation}
\tilde{z} = \frac{z}{H(t)}
\end{equation}
is a Lagrangian variable that follows the vertical expansion of the disc, $H(t)$ is the dynamical vertical scaleheight of the disc, and the quantities with tildes are normalized variables that satisfy a standard dimensionless form of the equations of vertical structure.

In order to preserve separability the (modified) Maxwell stress $M^{i j}$ must have the same vertical structure as the pressure ($\tilde{M} = \tilde{p}$)\footnote{We can have an additional height independent contribution to $M^{ij}$ (e.g.\ coming from a height-independent vertical magnetic field), but this has no effect on the dynamics.}. This assumption has a couple of important consequences. It corresponds to a plasma-$\beta$, defined as the ratio of hydrodynamic to magnetic pressure $\beta_m := p/p_m$, independent of height. Additionally it has implications for the realisability of $M^{ij}$: for a large scale field we require $M^{z z} = 0$ in order that the underlying magnetic field obeys the solenoidal condition. For small scale/turbulent fields the solenoidal condition instead implies the mean of $B^{z}$ is independent of height; however $M^{z z}$ has the same vertical structure as pressure.

The separated solution works provided that

\begin{equation}
\frac{d^2 H}{d t^2} = - \frac{G M_{\bullet}}{r^3} H + \frac{\hat{p}}{\hat{\rho} H} \left(1 + \frac{\hat{M}}{2 \hat{p}} - \frac{\hat{M}_{zz}}{\hat{p}} \right) ,
\end{equation}

\begin{equation}
\frac{d \hat{\rho}}{d t} = - \hat{\rho} \left ( \Delta + \frac{1}{H} \frac{d H}{d t} \right) ,
\end{equation}

\begin{equation}
\frac{d \hat{p}}{d t} = - \Gamma_1 \hat{p} \left ( \Delta + \frac{1}{H} \frac{d H}{d t} \right) + (\Gamma_3 - 1)\left(  f_{\mathcal{H}} n \hat{p}_v - \lambda \frac{\hat{F}}{H} \right) ,
\end{equation}

\begin{equation}
\hat{F} = \frac{16 \sigma \hat{T}^{4}}{3 \kappa \hat{\rho} H} \quad ,
\end{equation}

\begin{equation}
\hat{p} = (1 + \beta_r) \frac{\mathcal{R} \hat{\rho} \hat{T}}{\mu} \quad ,
\end{equation}
where the positive constant $\lambda$ is a dimensionless cooling rate that depends on the equations of vertical structure (further details can be found in Paper I) and 

\begin{equation}
\beta_r = \frac{4 \sigma \mu}{3 c \mathcal{R}}  \frac{\hat{T}^3}{\hat{\rho}} \quad .
\end{equation}
We must supplement these equations with a closure model for $M^{i j}$ and $f_{\mathcal{H}}$.

Note that the surface density and vertically integrated pressures are (owing to the definitions of the scaleheight and the dimensionless variables)

\begin{equation}
\Sigma = \hat{\rho} H , \quad P = \hat{p} H, \quad P_v = \hat{p}_v H .
\end{equation}
The vertically integrated heating and cooling rates are

\begin{equation}
f_{\mathcal{H}} n P_v ,\quad \lambda \hat{F} \quad .
\end{equation}
The cooling rate can also be written as

\begin{equation}
\lambda \hat{F} = 2 \sigma \hat{T}_s^4
\end{equation}
where $\hat{T}_{s} (t)$ is a representative surface temperature defined by

\begin{equation}
\hat{T}^{4}_s = \frac{8 \lambda}{3} \frac{\hat{T}^4}{\hat{\tau}}
\end{equation}
and

\begin{equation}
\hat{\tau} = \kappa \Sigma
\end{equation}
is a representative optical thickness.

We then have

\begin{equation}
\frac{1}{H} \frac{d^2 H}{d t^2} = - \frac{G M_{\bullet}}{r^3} + \frac{P}{\Sigma H^2} \Biggl(1 + \frac{\hat{M}}{2 p} - \frac{\hat{M}_{z z}}{p} \Biggr) ,
\end{equation}

\begin{equation}
J \Sigma = \mathrm{constant},
\end{equation}

\begin{equation}
\left( \frac{1}{\Gamma_3 - 1} \right) \frac{d P}{d t} = - \frac{\Gamma_1 P}{\Gamma_3 - 1} \left( \Delta + \frac{1}{H} \frac{d H}{d t} \right) + f_{\mathcal{H}} n P_v  - \lambda \hat{F} ,
\end{equation}
with

\begin{equation}
\frac{\lambda \hat{F}}{P n} = \lambda \frac{16 \sigma (\mu/\mathcal{R})^{4}}{3 \kappa n} P^{3} \Sigma^{- 5} (1 + \beta_r)^{-4} \quad .
\end{equation}

We assume for a given $\beta_m^{\circ}$, $\beta_r^{\circ}$ and $n$ there exists an equilibrium solution for a circular disc and use this solution to nondimensionalise the equations. As in the hydrodynamical models considered in Paper I, we use $^{\circ}$ to denote the equilibrium values in the reference circular disc (e.g. $H^{\circ}$, $T^{\circ}$ etc). Depending on the closure model there can be multiple equilibrium solutions, some of which can be unstable (particularly in the radiation dominated limit). Our choices of solution branch for our two closure models are specified in Appendices \ref{circular disc} and \ref{mag ref state}. 

Scaling $H$ by $H^{\circ}$, $\hat{T}$ by $T^{\circ}$, $M^{ij}$ by $p^{\circ}$, $t$ by $1/n$ and $J$ by $2/n$ we obtain the dimensionless version

\begin{equation}
\frac{\ddot{H}}{H} = -(1 - e \cos E)^{-3} + \frac{T}{H^2} \frac{1 + \beta_r}{1 + \beta_{r}^{\circ} } \frac{ \Biggl( 1 + \frac{1}{2} \frac{M}{p} - \frac{M^{z z}}{p} \Biggr)}{\left[ 1 + \frac{1}{2} \frac{M^{\circ}}{p^{\circ}} - \frac{(M^{z z})^{\circ}}{p^{\circ}} \right]} ,
\label{scale height equation}
\end{equation}

\begin{align}
\begin{split}
 \dot{T} &= - (\Gamma_3 - 1) T \left(\frac{\dot{J}}{J} + \frac{\dot{H}}{H} \right) \\
 &+ (\Gamma_3 - 1) \frac{1 + \beta_r}{1 + 4 \beta_r} T \left( f_{\mathcal{H}} \frac{P_v}{P} - \mathcal{C}^{\circ}  \frac{1 + \beta_{r}^{\circ} }{1 + \beta_r} J^2 T^{3} \right) ,
 \end{split}
 \label{thermal energy equation T}
\end{align}
where a dot over a letter indicates a derivative with respect to rescaled time. We have written the thermal energy equation in terms of the temperature. The factor $\frac{\Gamma_{3} - 1}{1 + 4 \beta} \propto \frac{1}{c_{V}}$ where $c_{V}$ is the specific heat capacity at constant volume. $\beta_r$ can be obtained through

\begin{equation}
\beta_r = \beta_{r}^{\circ}  J H T^3 \quad ,
\end{equation}
where we have introduced $\beta_{r}^{\circ} $, which is the $\beta_{r}$ of the reference circular disc. The equilibrium values of the reference circular disc $H^{\circ}$, $\hat{T}^{\circ}$, etc., are determined by $\beta_r^{\circ}$ and $n$. The reference cooling rate can be obtained by setting it equal to the reference heating rate: $\mathcal{C}^{\circ} = f_{\mathcal{H}}^{\circ} \frac{P_{v}^{\circ}}{P^{\circ}}$.

Additionally we introduce the (nondimensional) entropy,

\begin{equation}
s := 4 \beta_r + \ln (J H T^{1/(\gamma - 1)}) \quad ,
\end{equation}
which has contributions from the radiation and the gas.

\section{Effect of Magnetic Fields} \label{magnetic pressure effect}

\subsection{Magnetic fields in eccentric discs}

In Paper I we found that (when $p_v = p_g$) our radiation dominated solutions exhibit extreme compression at pericentre, similar to the extreme behaviour of the adiabatic solutions of \citet{Ogilvie14}. Many of our solutions with more moderate behaviour have strong viscous stresses at pericentre which call into question the validity of the $\alpha-$prescription. 

What additional physical processes could reverse the collapse of the fluid column and prevent the extreme compression seen in the radiation dominated model? Can the collapse be reversed without encountering unphysically strong viscous stresses? An obvious possibility is the presence of a large scale horizontal magnetic field within the disc which will resist vertical compression. Such a field could be weak for the majority of the orbit but, owing to the relative incompressibility of magnetic fields, become dynamically important during the maximum compression at pericentre. In Appendix \ref{mag deriv} we show that in an eccentric disc, a solution to the steady ideal induction equation in an inertial frame is

\begin{equation}
B^{a} = 0 ,\quad B^{E} = \frac{\Omega B^{E}_0 (a,\tilde{z})}{n J H} ,\quad B^{z}  = \frac{B^{z}_0 (a)}{J} \quad .
\end{equation}
Here $B^{E}$ is the component parallel to the orbital motion (quasi-toroidal) and $B^{z}$ is the vertical component. We use quasi-poloidal to indicate the components $B^{a}$ and $B^{z}$.

The magnetic field of a star undergoing tidal disruption has been studied by \citet{Guillochon17} and \citet{Bonnerot17}. In these papers it was found that the stretching of the fields during the disruption causes an increase in the magnetic pressure from the field aligned with the orbital direction. Meanwhile the gas pressure and magnetic pressure from the field perpendicular to the orbit drop. \citet{Guillochon17} found that this tends to result in the magnetic pressure from the parallel field becoming comparable to the gas pressure. Similar results were found in \citet{Bonnerot17}, although with a dependence on the initial field direction. This supports our adopted field configuration, with the vertical field set to zero. As the vertical field does not contribute to the dynamics of the vertical oscillator we can do so without loss of generality.

In addition to the large scale magnetic field, we assume that the effects of the small-scale/turbulent magnetic field can be modelled by an $\alpha$-viscosity, 

\begin{equation}
\mu_{s,b} = \alpha_{s,b} \frac{p_v}{\omega_{\rm orb}} ,
\end{equation}
where $\mu_{s,b}$ are the dynamic shear and bulk viscosities, $\alpha_{s,b}$ are dimensionless coefficients, $\omega_{\rm orb}$ is some characteristic frequency of the orbital motion (here taken to be $n$) and $p_v$ is some choice of pressure. As in Paper I we set the bulk viscosity to zero ($\alpha_b=0$).

As discussed in Section \ref{breathinmode}, in order to preserve separability of the equations we require $B^{E}_0 (a,\tilde{z})$ to depend on $\tilde{z}$ in such a way as to make $\beta_m$ independent of height. The dimensionless equations for the variation of the dimensionless scale height $H$ and temperature  $T$ around the orbit (derived in Appendix \ref{mag deriv}) are then

\begin{align}
\begin{split}
 \frac{\ddot{H}}{H} &= -(1 - e \cos E)^{-3} + \frac{T}{ H^2} \frac{1 + \beta_r}{1 + \beta_{r}^{\circ} } \left(1 + \frac{1}{\beta_m^{\circ}} \right)^{-1} \\
 &\times \Biggl[1 + \frac{1}{\beta_m} - 2 \alpha_s \frac{P_v}{P} \frac{\dot{H}}{H} - \left(\alpha_b - \frac{2}{3} \alpha_s \right) \frac{P_v}{P} \left(\frac{\dot{J}}{J} + \frac{\dot{H}}{H}\right) \Biggr] ,
 \end{split}
\end{align}

\begin{align}
\begin{split}
 \dot{T} &= - (\Gamma_3 - 1) T \left(\frac{\dot{J}}{J} + \frac{\dot{H}}{H} \right) \\
&+ (\Gamma_3 - 1) \frac{1 + \beta_r}{1 + 4 \beta_r} T \left( f_{\mathcal{H}} \frac{P_v}{P}  - \frac{9}{4} \alpha_s \frac{P_v^{\circ}}{P^{\circ}} \frac{1 + \beta_{r}^{\circ} }{1 + \beta_r} J^2 T^{3} \right) ,
\end{split}
\end{align} 
and the plasma-$\beta$ is given by

\begin{equation}
\beta_m = \beta_{m}^{\circ}  J H T \frac{1 + \beta_r}{1 + \beta_{r}^{\circ} } \frac{1 - e \cos E}{1 + e \cos E}
\end{equation}
where $\beta_{m}^{\circ}$ is the plasma beta in the reference circular disc.  

These equations can be solved using the same relaxation method used to solve the purely hydrodynamic equations in Paper I. However caution must be taken when solving the equations with low $\beta_{m}^{\circ} $ (i.e. strong magnetic fields throughout the disc) as the method does not always converge to a periodic solution (or at least takes an excessively long time to do so). This is most likely due to the absence of dissipative effects acting on the magnetic field, so any free oscillations in the magnetic field are not easily damped out. We believe that the quasiperiodic solutions we find for low $\beta_{m}^{\circ} $ are the superposition of the forced solution and a free fast magnetosonic mode. For now we only consider values of $\beta_{m}^{\circ} $ which successfully converge to a periodic solution.

\subsection{Viscous stress independent of the magnetic field ($p_v = p_g$)}

\begin{figure}
\centering
\includegraphics[trim=0 0 0 0,clip,width=0.9\linewidth]{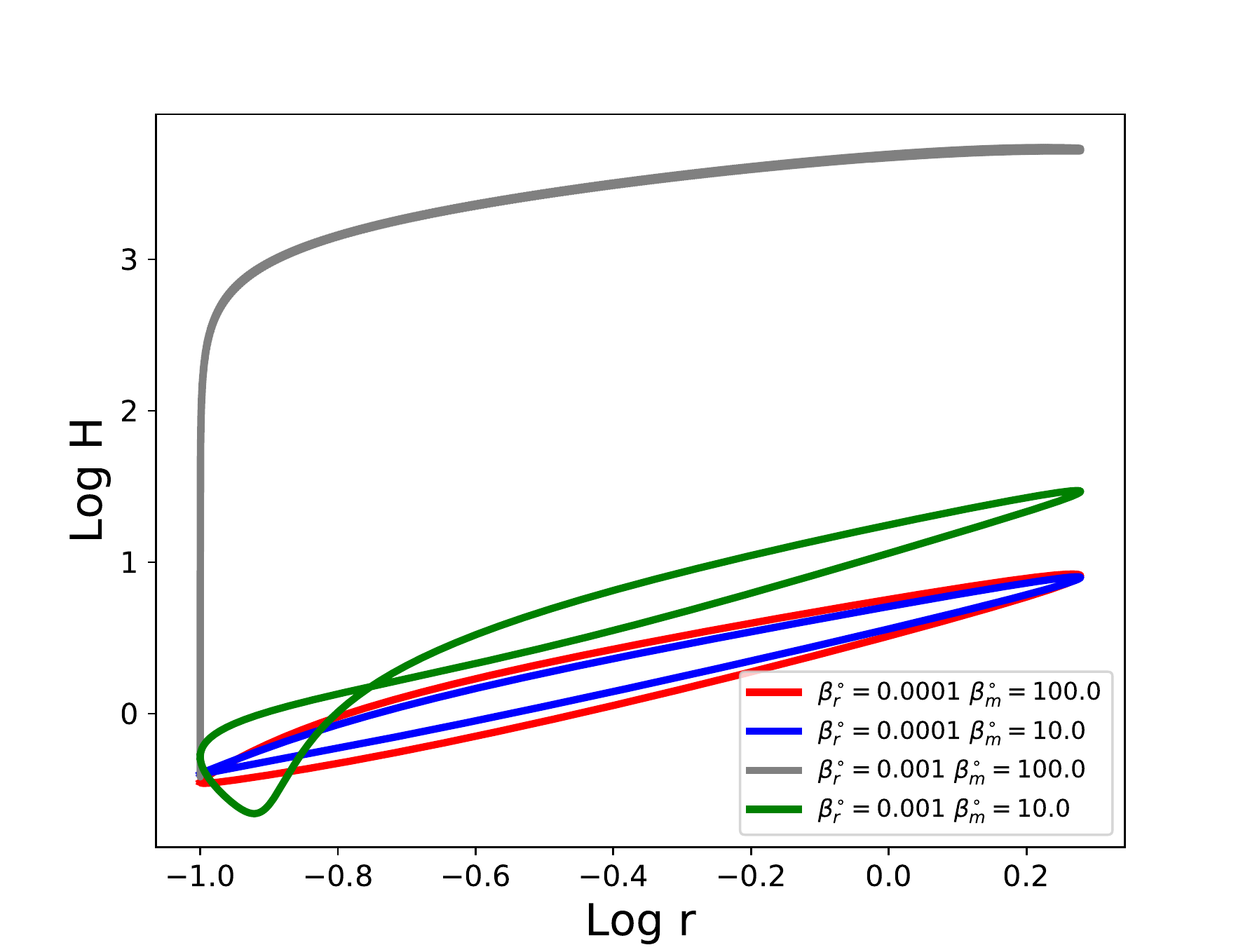}
\caption{Variation of the scale height of the disc with radiation + gas pressure with different $\beta_r^{\circ}$ and magnetic fields. Disc parameters are $p_v = p_g$, $\alpha_s=0.1$, $\alpha_b=0$, $e=q=0.9$ and $E_0 = 0$. Red line has $\beta_{r}^{\circ}  = 10^{-4}$ and $\beta_{m}^{\circ} = 100$, blue line has $\beta_{r}^{\circ}  = 10^{-4}$ and $\beta_{m}^{\circ} = 10$, grey line has $\beta_{r}^{\circ}  = 10^{-3}$ and $\beta_{m}^{\circ} = 100$, green line has $\beta_{r}^{\circ}  = 10^{-3}$ and $\beta_{m}^{\circ} = 10$. The discs with $\beta_{m}^{\circ}=100$ are nearly indistinguishable from an unmagnetised disc.} 
\label{magnetic comparison}
\end{figure}

\begin{figure}
\centering
\includegraphics[trim=0 0 0 0,clip,width=0.9\linewidth]{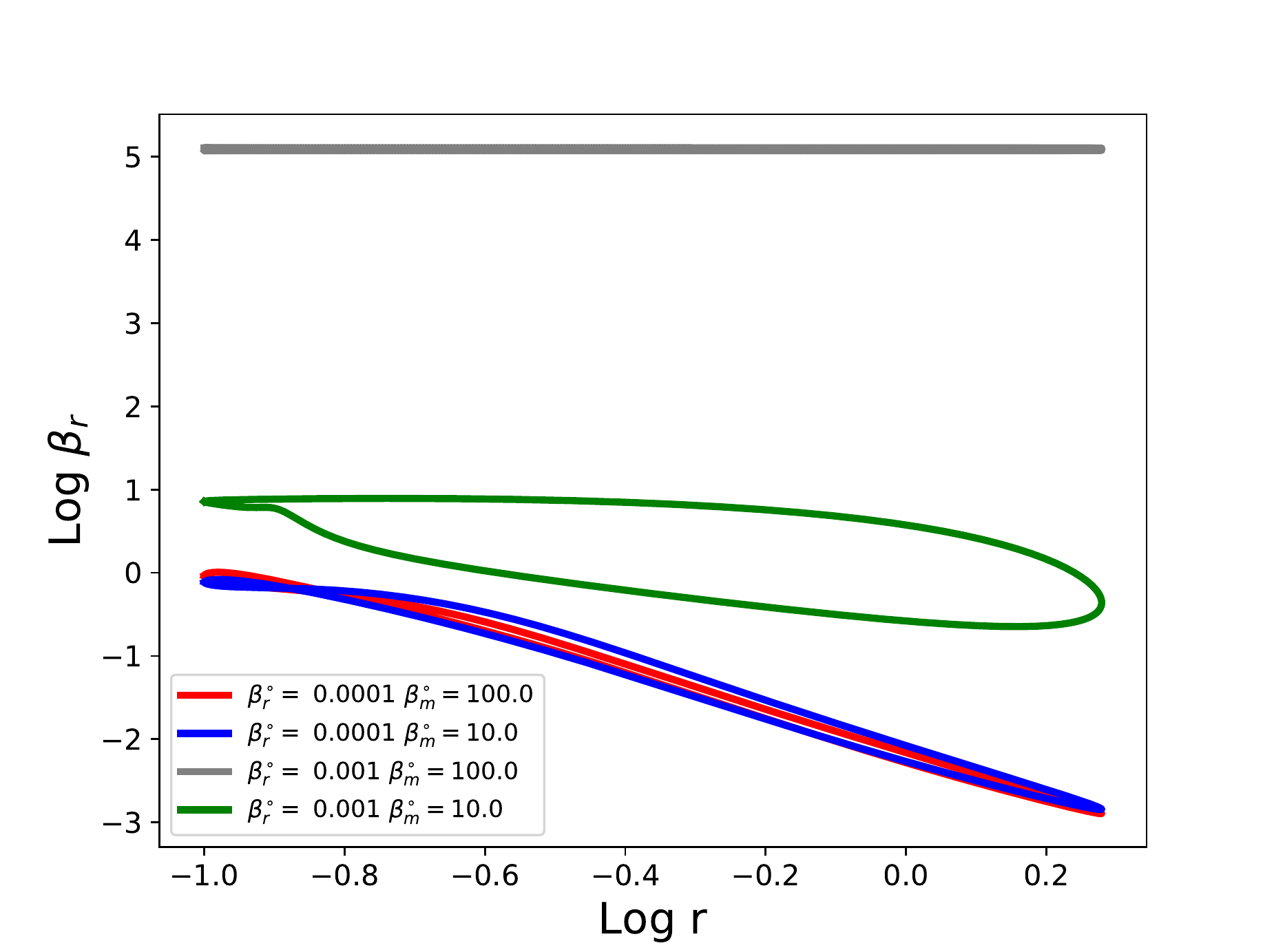}
\caption{Variation of the ratio of radiation to gas pressure ($\beta_r$) around around the orbit for each model in Figure \ref{magnetic comparison}.} 
\label{magnetic comparison betaplot}
\end{figure}

\begin{figure}
\centering
\includegraphics[trim=0 0 0 0,clip,width=0.9\linewidth]{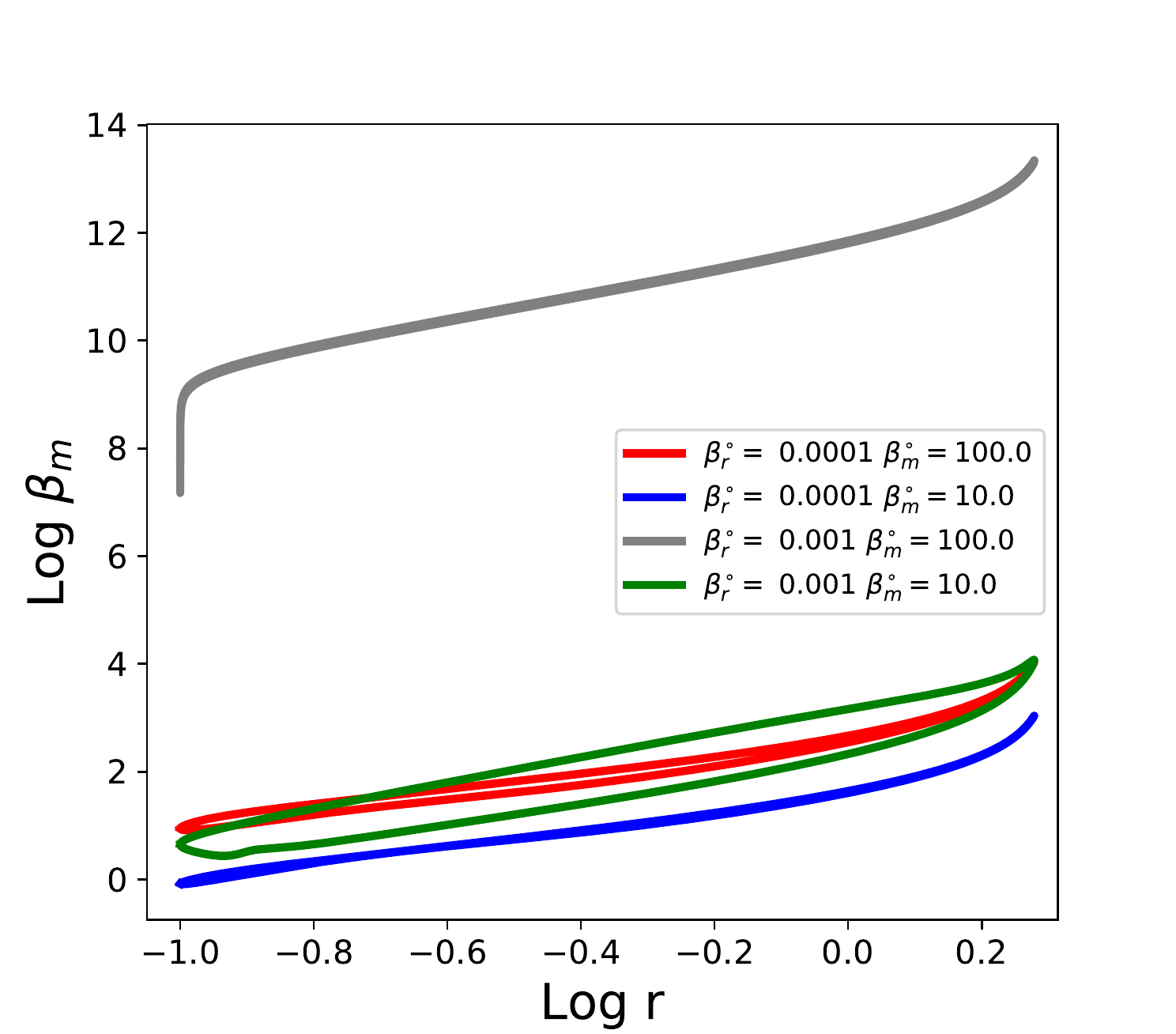}
\caption{Variation of the plasma-$\beta$ around the orbit for each model in Figure \ref{magnetic comparison}.} 
\label{magnetic comparison plasmabetaplot}
\end{figure}

Figures \ref{magnetic comparison}-\ref{magnetic comparison plasmabetaplot} show the variations of the scale height, $\beta_r$ and $\beta_m$ around the orbit for a disc with $\alpha_s=0.1$, $\alpha_b = 0$, $e=q=0.9$ and $E_0 = 0$. The magnetic field has a weak effect on the gas pressure dominated ($\beta_{r}^{\circ} = 10^{-4}$) solutions. For the radiation pressure dominated ($\beta_{r}^{\circ} = 10^{-3}$) case, a strong enough magnetic field stabilises the solution against the thermal instability and, instead of the nearly adiabatic radiation dominated solutions seen in the hydrodynamic case, the solution is only moderately radiation pressure dominated and maintains significant entropy variation around the orbit. This solution is similar to the moderately radiation pressure dominated hydrodynamic solutions. If the field is too weak (e.g. $\beta_{m}^{\circ} = 100$ considered here) the magnetic field isn't capable of stabilising against the thermal instability and the solution tends to the nearly adiabatic radiation dominated solution.

Most of the solutions in Figures \ref{magnetic comparison}-\ref{magnetic comparison plasmabetaplot} are not sufficiently radiation pressure dominated to represent most TDEs. Figures \ref{magnetic highbeta  comparison}-\ref{magnetic highbeta  comparison plasmabetaplot} show solutions which attain much higher $\beta_r$. We see it is possible to attain significantly radiation pressure dominated solutions which do not possess the extreme variation of the scale height around the orbit present in the hydrodynamic case. In particular, consider the green curve with $\beta_{r}^{\circ} = 1$, $\beta_m^{\circ} = 0.005$. Like the radiation dominated hydrodynamic solutions the solution with $\beta_{r}^{\circ} = 1$, $\beta_m^{\circ} = 0.005$ is nearly adiabatic; however magnetic pressure dominates over radiation pressure during pericentre passage. This additional support at pericentre prevents the extreme compression, and resultant heating, seen in the hydrodynamic model - resulting in more moderate variation of the scale height around the orbit. Unlike the similarly radiation dominated, unmagnetised, solutions considered in Paper I, this solution remains consistent with the thin disc assumptions for typical TDE parameters. 

It should be cautioned that the grey solution (with $\beta_r^{\circ}=1$, $\beta_m^{\circ}=1$) in Figures \ref{magnetic highbeta  comparison}-\ref{magnetic highbeta  comparison plasmabetaplot} has not converged. The magnetic field is unimportant for this solution. Based on the radiation dominated hydrodynamic models of Paper I, the disc with $\beta_r^{\circ}=1$ will converge on a solution with $\beta_r$ much larger than the $\beta_r \sim 10^5-10^6$ which were the most radiation dominated, converged, solutions obtained in Paper I. As the entropy gained per orbit is small compared to the entropy in the disc, this will take a large number of orbits ($> 10 000$ orbits) to converge, so the converged solution isn't of much interest when considering transient phenomena like a TDE.

\begin{figure}
\centering
\includegraphics[trim=0 0 0 0,clip,width=0.9\linewidth]{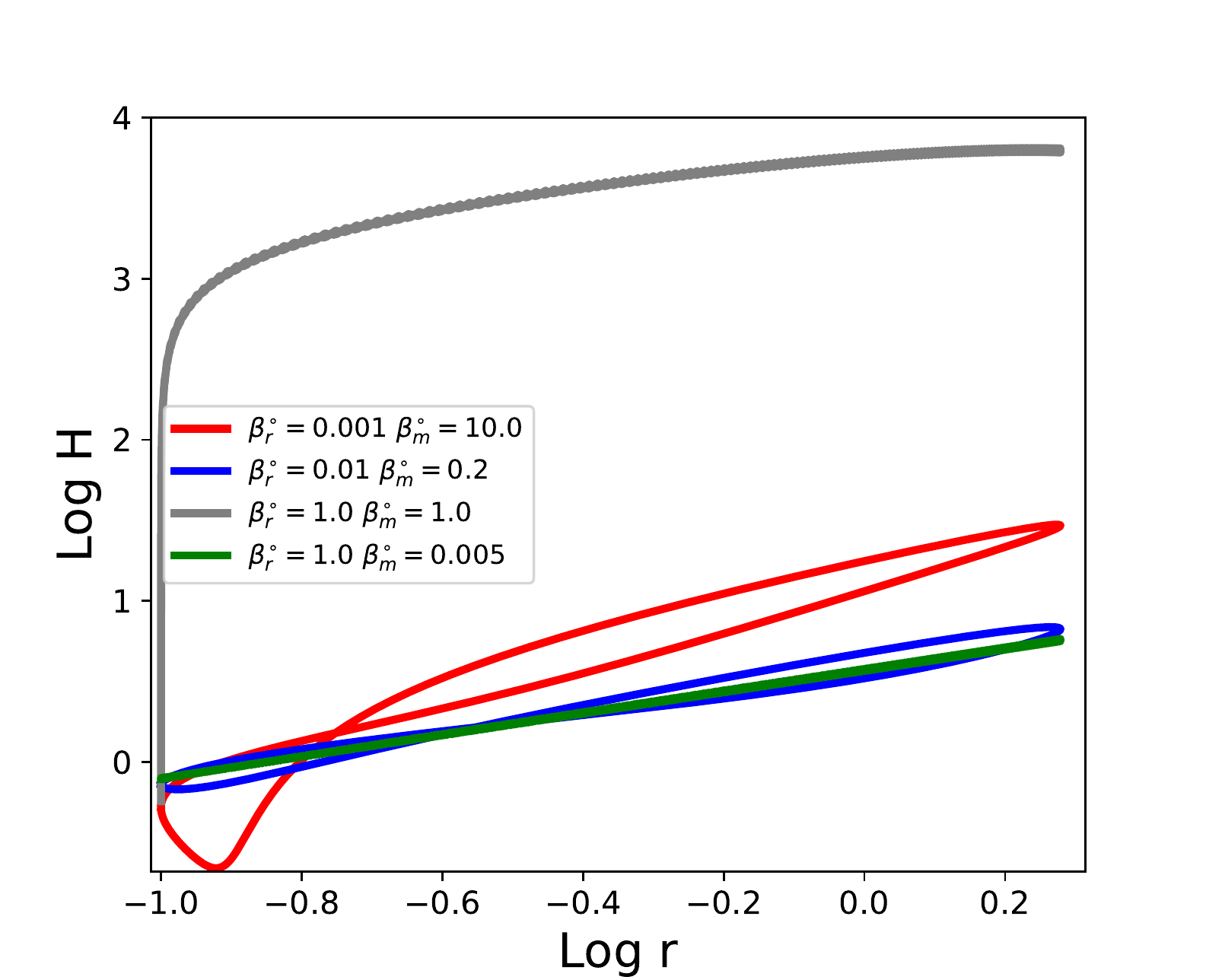}
\caption{Same as Figure \ref{magnetic comparison} but attaining larger $\beta_r$. Red line has $\beta_{r}^{\circ}  = 10^{-3}$ and $\beta_{m}^{\circ} = 10$, blue line has $\beta_{r}^{\circ}  = 10^{-2}$ and $\beta_{m}^{\circ} = 0.2$, grey line has $\beta_{r}^{\circ} = 1$ and $\beta_{m}^{\circ} = 1$, green line has $\beta_{r}^{\circ}  = 1$ and $\beta_{m}^{\circ} = 0.005$.} 
\label{magnetic highbeta comparison}
\end{figure}

\begin{figure}
\centering
\includegraphics[trim=0 0 0 0,clip,width=0.9\linewidth]{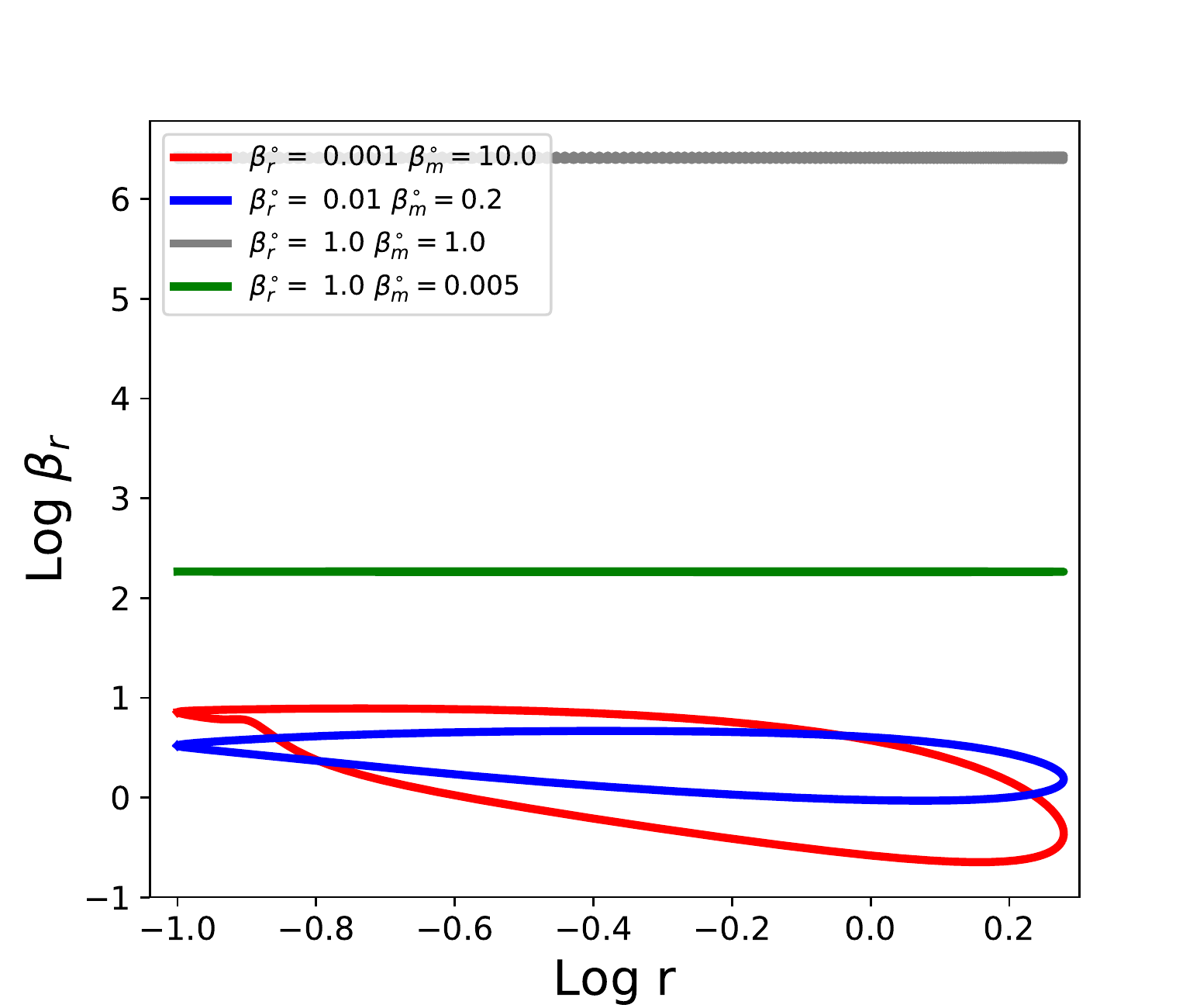}
\caption{Variation of $\beta_r$ around the orbit for each model in Figure \ref{magnetic highbeta comparison}.} 
\label{magnetic highbeta comparison betaplot}
\end{figure}

\begin{figure}
\centering
\includegraphics[trim=0 0 0 0,clip,width=0.9\linewidth]{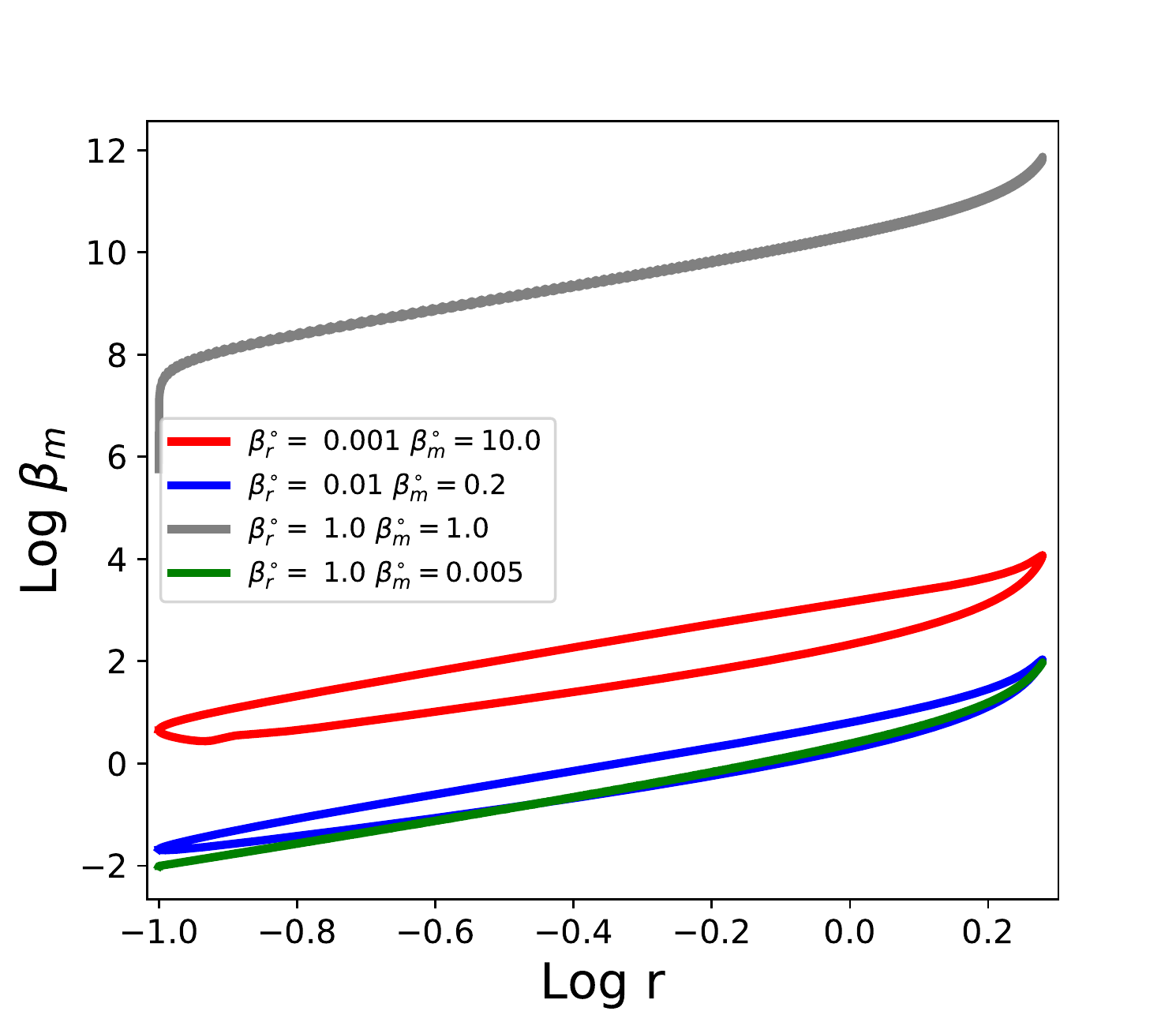}
\caption{Variation of the plasma-$\beta$ around the orbit for each model in Figure \ref{magnetic highbeta comparison}.} 
\label{magnetic highbeta comparison plasmabetaplot}
\end{figure}
 
\begin{figure}
\includegraphics[trim=10 0 10 10,clip,width=\linewidth]{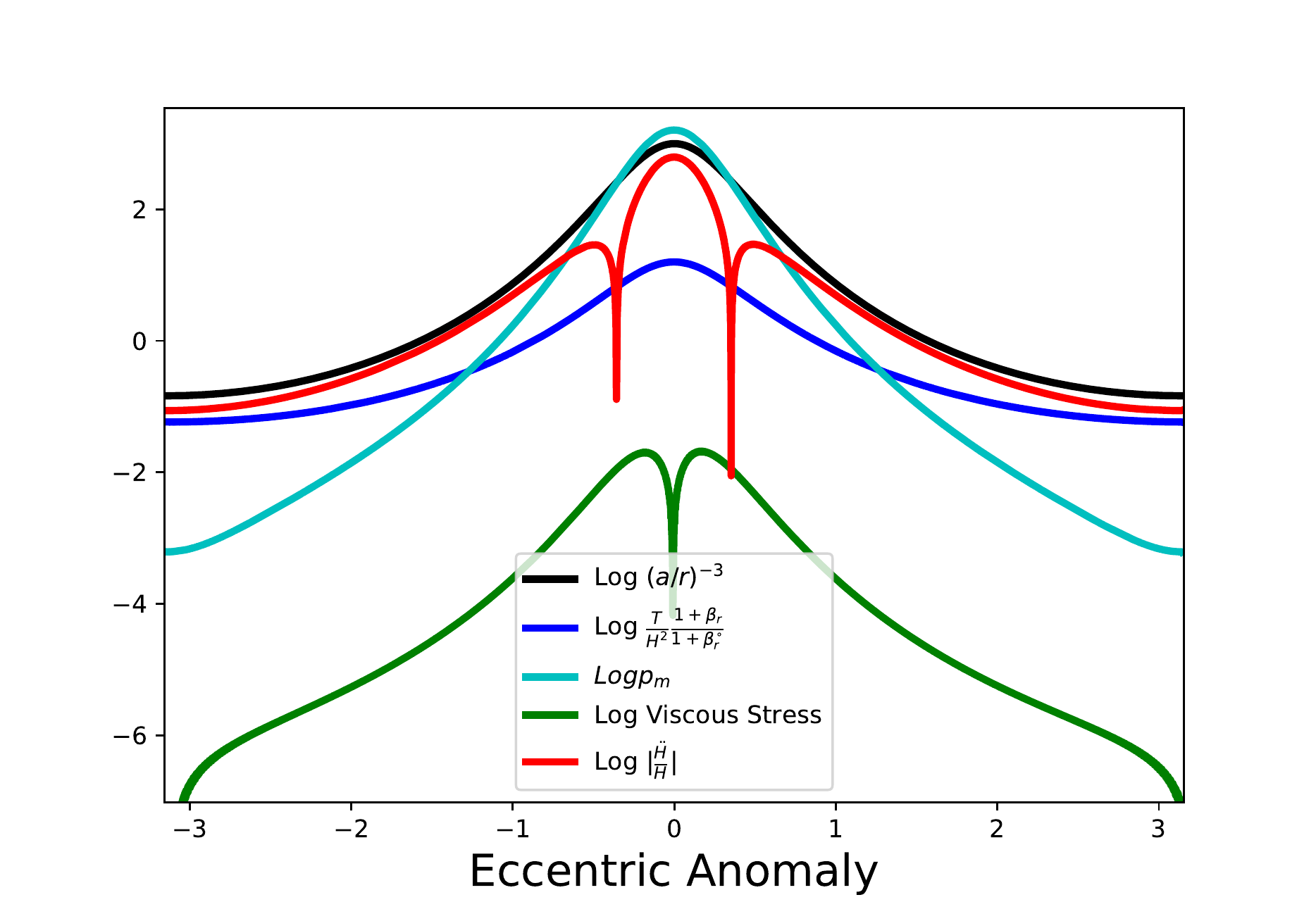}
\caption{Magnitude of the different terms in the vertical momentum equation for a magnetised radiation-gas mixture; the black line is the disc gravity, the blue is the hydrodynamic pressure, the cyan is the magnetic pressure, green is the viscous stress and red is the vertical momentum. Disc parameters are $p_v = p_g$, $\beta_{r}^{\circ} = 1$, $\beta_m^{\circ} = 0.005$, $\alpha_s=0.1$, $\alpha_b=0$, $e=q=0.9$ and $E_0 = 0$. The balance at pericentre is now between the gravity, the magnetic pressure and the vertical acceleration.} 
\label{mag forcebalance}
\end{figure}

 Figure \ref{mag forcebalance} shows the magnitude of different terms in the momentum equation for a disc with $p_v = p_g$, $\beta_{r}^{\circ} = 1$, $\beta_m^{\circ} = 0.005$, $\alpha_s=0.1$, $\alpha_b=0$, $e=q=0.9$ and $E_0 = 0$ (i.e. the green solution from Figures \ref{magnetic highbeta  comparison}-\ref{magnetic highbeta  comparison plasmabetaplot}). This shows that the dominant balance in this solution is between the vertical acceleration, gravity and the magnetic force. This suggests that dynamics of radiation pressure dominated TDEs may be controlled by the magnetic field. Being the least compressible pressure term, the magnetic pressure tends to dominate at pericentre, even if it is fairly weak throughout the rest of the disc. However for radiation dominated TDEs pressure is only important near pericentre so even a weak magnetic field will have a disproportionate contribution to the dynamics. This suggests that ignoring even subdominant magnetic fields in TDE discs can lead to fundamental changes to the TDE dynamics.  
 
While it is possible to find a combination of $\beta_r^{\circ}$ and $\beta_m^{\circ}$ which yields a solution with the desired $\beta_r$ exhibiting ``reasonable'' behaviour, it is not clear that the magnetic field in the disc will always be strong enough to produce the desired behaviour. It is possible that this represents a tuning problem for $\beta_m^{\circ}$. 

\begin{figure}
\includegraphics[trim=0 0 0 0,clip,width=\linewidth]{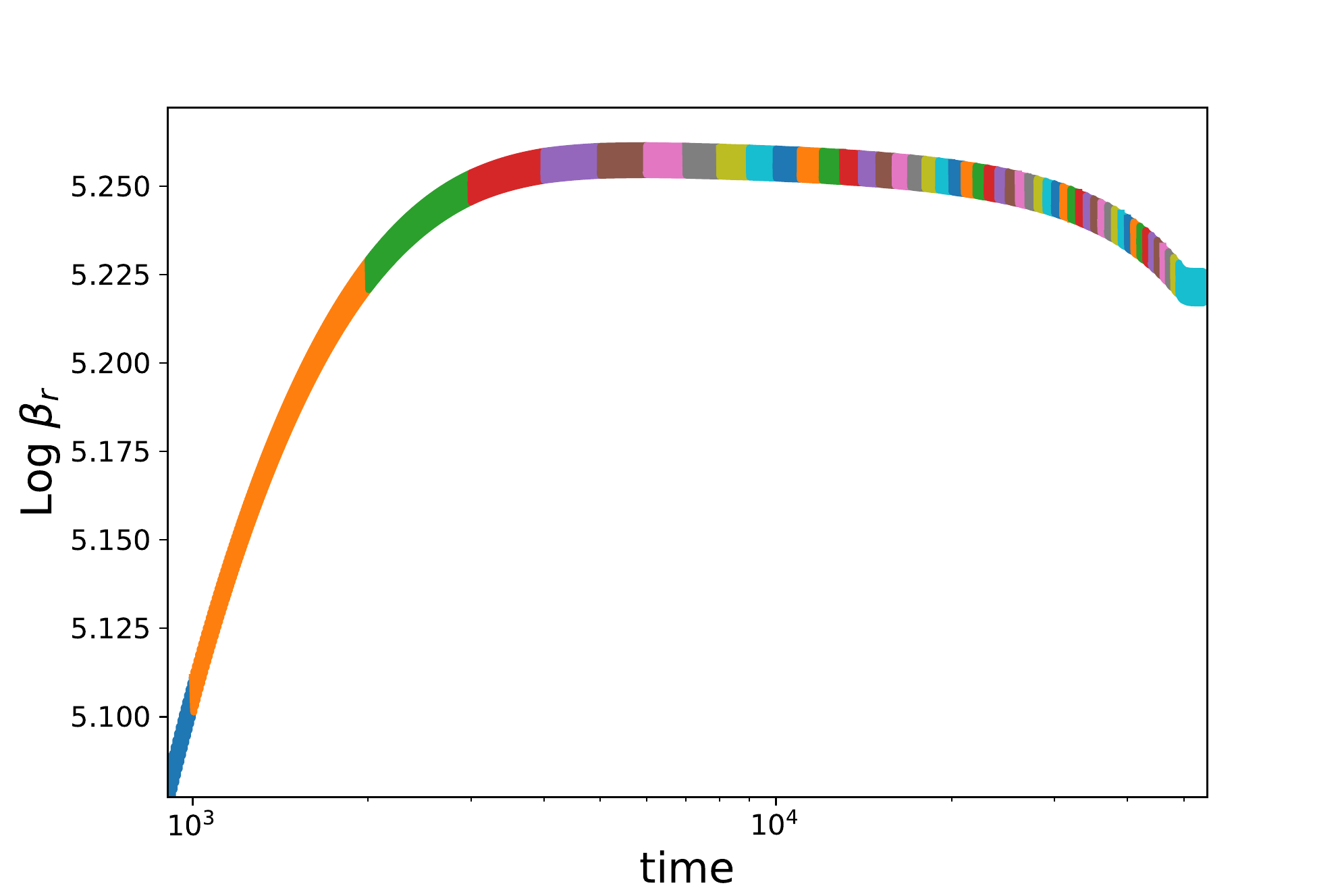}
\caption{$\beta_r$ for a disc where the magnetic field strength is gradually increased from $\beta_m^{\circ} = 100$ to $\beta_m^{\circ} = 10$. Disc parameters are $p_v = p_g$, $\beta_r^{\circ} = 10^{-3}$, $\alpha_s=0.1$, $\alpha_b=0$, $e=q=0.9$ and $E_0 = 0$. Colours indicate where we have stopped the calculation and restarted with a different magnetic field strength. The resulting solution remains in the nearly adiabatic radiation pressure dominated state and doesn't converge on the green solution of Figures \ref{magnetic comparison}-\ref{magnetic comparison plasmabetaplot}. The solution with the final magnetic field strength ($\beta_m^{\circ} = 10$) was run for longer to allow it to relax to a steady state.} 
\label{mag field growth}
\end{figure}

To explore this we look at what happens if $\beta_m^{\circ}$ is initially too weak to stabilise against the thermal instability but we gradually raise it over several thermal times. Figure \ref{mag field growth} shows what happens when the magnetic field is increased gradually from $\beta_m^{\circ} = 100$ to $\beta_m^{\circ} = 10$ for a disc with $p_v = p_g$, $\beta_r^{\circ} = 10^{-3}$, $\alpha_s=0.1$, $\alpha_b=0$, $e=q=0.9$ and $E_0 = 0$. This corresponds to moving from the grey to the green solution in Figures \ref{magnetic comparison}-\ref{magnetic comparison plasmabetaplot}. This is done by periodically stopping the calculation and restarting with a larger $\beta_m^{\circ}$ The resulting $\beta_r$ in fact increases with time and remains close to that of the grey solution in Figures \ref{magnetic comparison}-\ref{magnetic comparison plasmabetaplot} even as we increase the magnetic field strength, and does not transition to a value consistent with the green solution. This suggests that the solution is sensitive to the path taken and that a magnetic field which grows (from an initially weak seed field), in a nearly adiabatic radiation pressure dominated disc, may not cause the disc to collapse to the gas pressure dominated branch. This is likely because the disc is very expanded meaning the magnetic field is still quite weak and incapable of influencing the dynamics. 

We carried out a similar calculation for a disc with $p_v = p_g$, $\beta_r^{\circ} = 1$, $\alpha_s=0.1$, $\alpha_b=0$, $e=q=0.9$ and $E_0 = 0$ moving from $\beta_m^{\circ} = 1$ to $\beta_{m}^{\circ} = 5 \times 10^{-3}$ (corresponding to the grey and green solutions of Figures \ref{magnetic highbeta comparison}-\ref{magnetic highbeta comparison plasmabetaplot}). In this case $\beta_r$ steadily increases with time (apart from a small variation over the orbital period) with the magnetic field having no appreciable influence on the solution. Owing to the relatively large $\beta_r^{\circ}$ this solution never reached steady state, as discussed previously. The implication of these two tests is that radiation pressure dominated, magnetised, discs can have two stable solution branches, with the choice of branch determined by the magnetic field history.

\begin{figure}
\includegraphics[trim=0 0 0 0,clip,width=\linewidth]{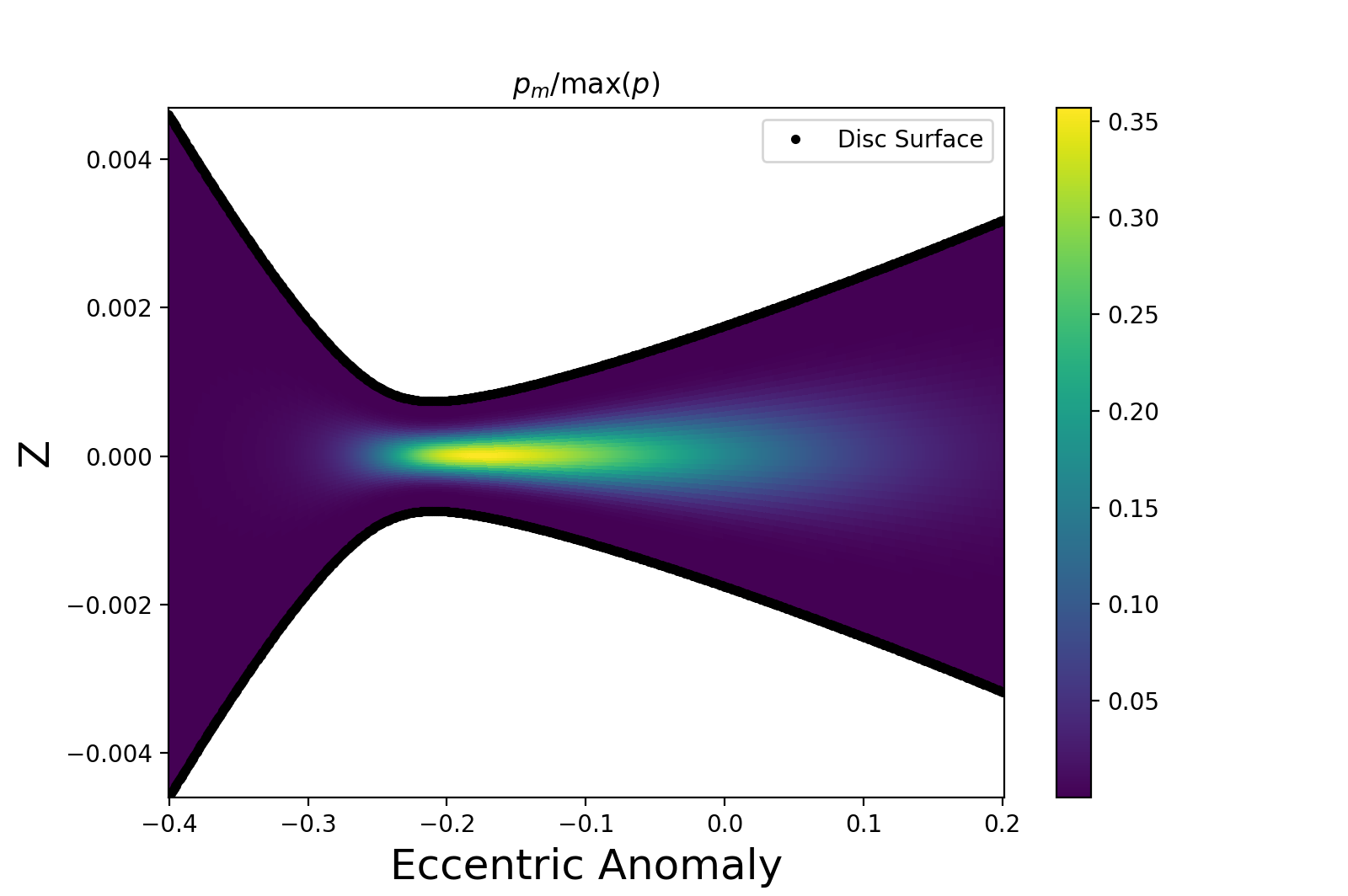}
\caption{Pericentre passage for a magnetised disc with $p_v = p_g$, $\beta_{r}^{\circ} = 10^{-3}$, $\beta_m^{\circ} = 10$, $\alpha_s=0.1$, $\alpha_b=0$, $e=q=0.9$ and $E_0 = 0$ showing the magnetic pressure scaled by the maximum hydrodynamic pressure. The magnetic field is highly concentrated in the nozzle. An unmagnetised disc, with the same parameters, is thermally unstable and would be considerably thicker.}  
\label{mag nozzle}
\end{figure}

Figure \ref{mag nozzle} shows the pericentre passage for a magnetised disc with $p_v = p_g$, $\beta_{r}^{\circ} = 10^{-3}$, $\beta_m^{\circ} = 10$, $\alpha_s=0.1$, $\alpha_b=0$, $e=q=0.9$ and $E_0 = 0$. The magnetic pressure is extremely concentrated within the nozzle and near to the midplane. Like the hydrodynamic nozzle structure considered in Paper I, the nozzle is asymmetric and located prior to pericentre, which is appears to be characteristic of dissipative highly eccentric discs.

\subsection{Viscous stress dependent on the magnetic field} \label{mag dependant stress}

\citet{Begelman07} have suggested that discs with strong toroidal fields may be stable to the thermal instability if the stress depends on the magnetic pressure. In this subsection we explore this possibility for a highly eccentric disc.

\begin{figure}
\centering
\includegraphics[trim=0 0 0 0,clip,width=0.9\linewidth]{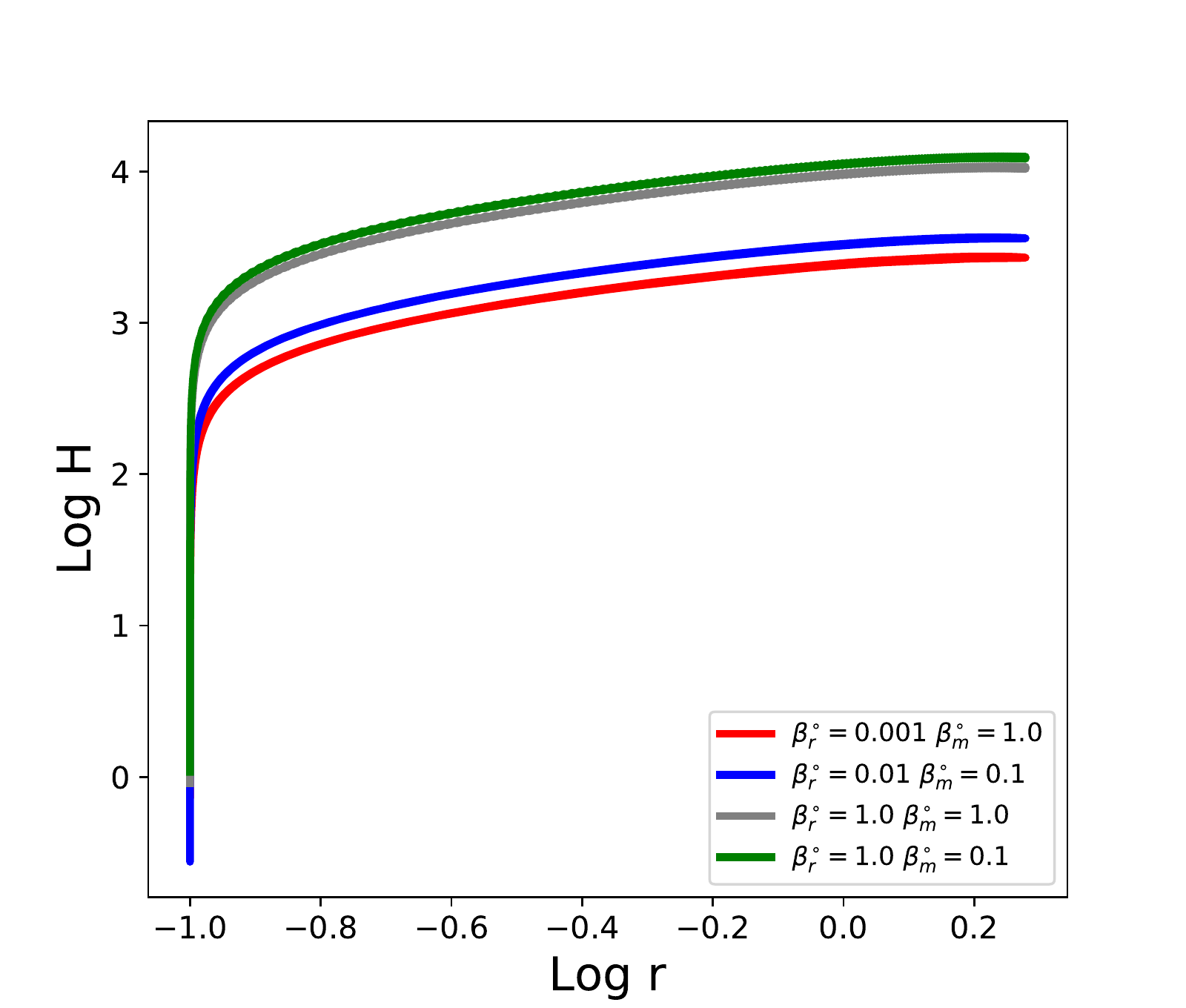}
\caption{Variation of the scale height of the disc when the viscous stress is proportional to the magnetic pressure. Disc parameters are $p_v = p_m$, $\alpha_s=0.1$, $\alpha_b=$, $e=q=0.9$ and $E_0 = 0$.} 
\label{pv eq pm}
\end{figure}

\begin{figure}
\includegraphics[trim=0 0 0 0,clip,width=0.9\linewidth]{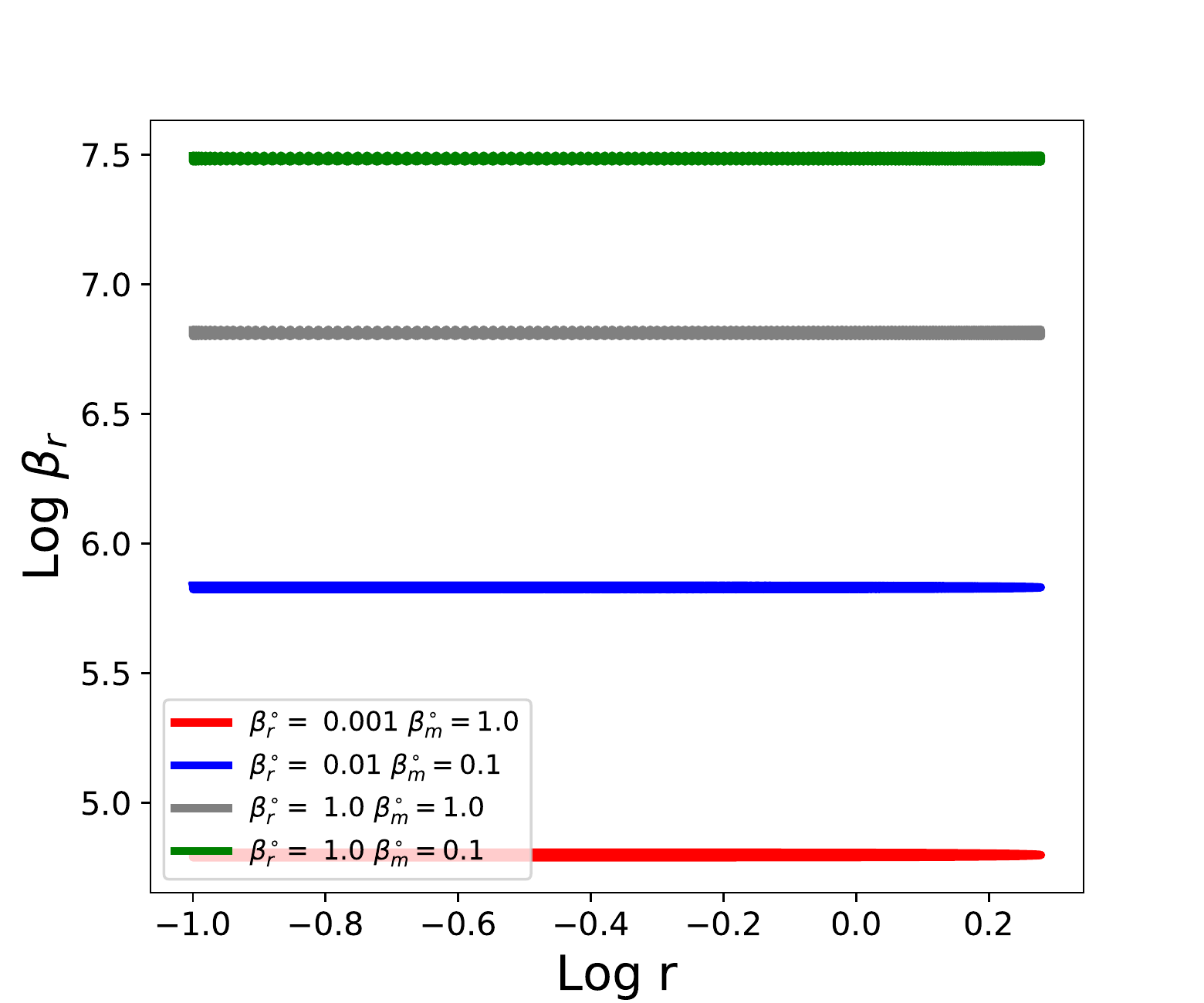}
\caption{Variation of $\beta_r$ around the orbit for each model in Figure \ref{pv eq pm}. }
\label{pv eq pm betar}
\end{figure}

\begin{figure}
\centering
\includegraphics[trim=0 0 0 0,clip,width=0.9\linewidth]{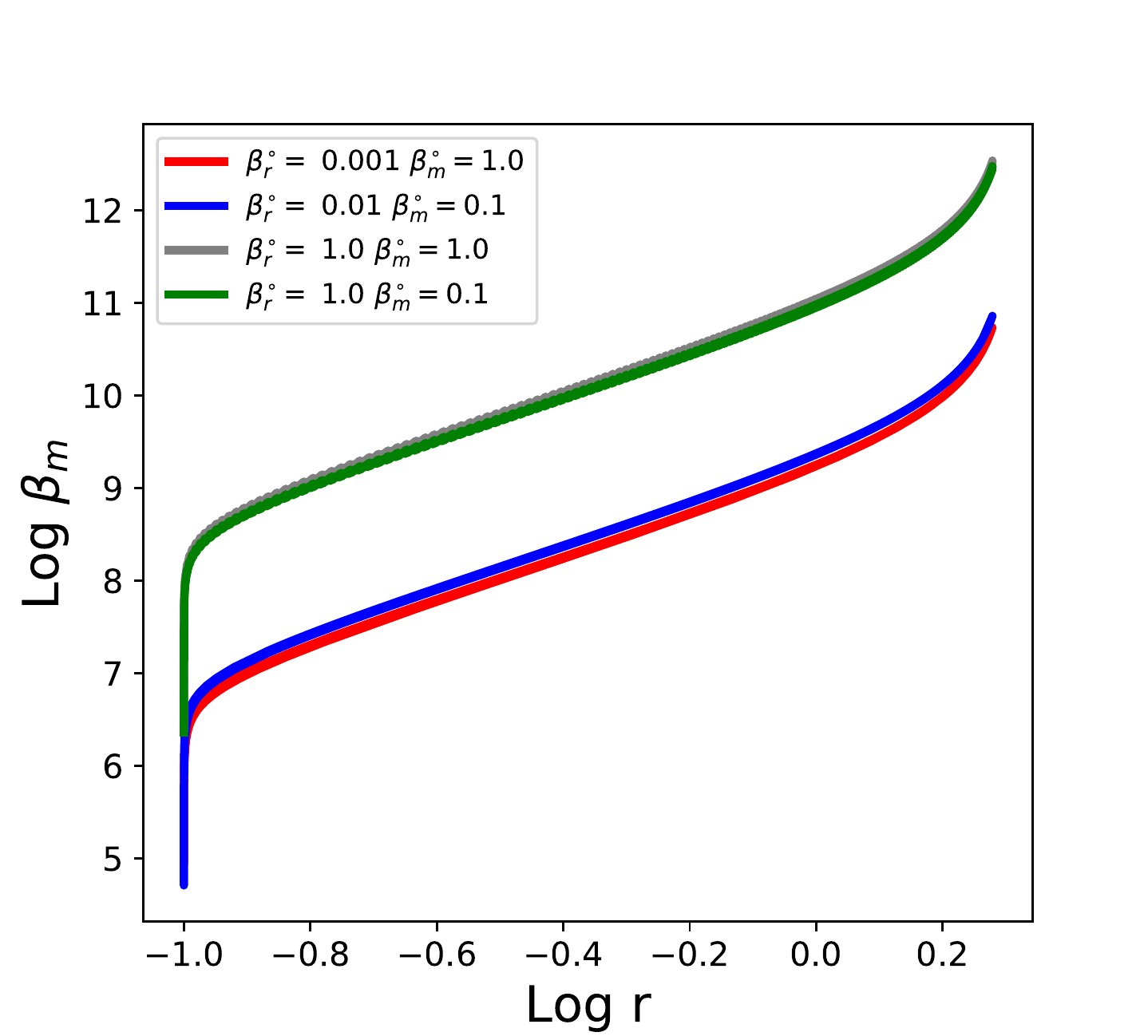}
\caption{Variation of the plasma-$\beta$ around the orbit for each model in Figure \ref{pv eq pm}.} 
\label{pv eq pm betam}
\end{figure}

Figures \ref{pv eq pm}-\ref{pv eq pm betam} show variation of the scale height, $\beta_r$ and plasma beta for a disc with $p_v = p_m$, $\alpha_s=0.1$, $\alpha_b=0$, $e=q=0.9$ and $E_0 = 0$. These have essentially the same behaviour as the nearly adiabatic radiation pressure dominated discs for the hydrodynamic case. This is not surprising as in this limit the gas and magnetic pressures are essentially negligible, which also results in negligible viscous stress/heating when it scales with either of these pressures. Increasing the magnetic field strength stabilises the ``gas pressure dominated" branch, where the magnetic field and viscous dissipation become important. This branch can have $p_r \gg p_g$ around the entire orbit; this is similar to the behaviour of the radiation pressure dominated hydrodynamic discs considered in Paper I with large $\alpha_s$.

\begin{figure}
\centering
\includegraphics[trim=0 0 0 0,clip,width=0.9\linewidth]{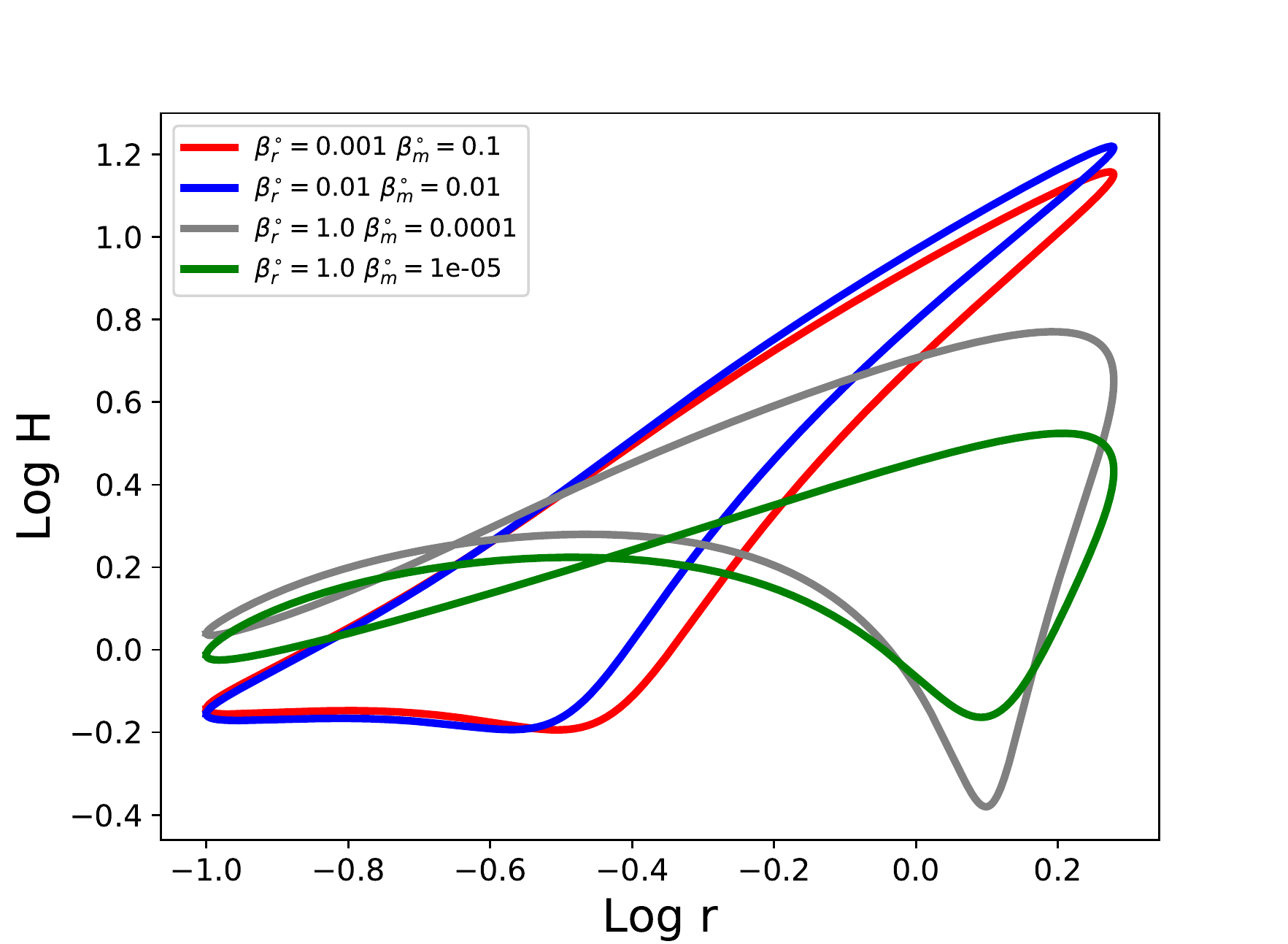}
\caption{Variation of the scale height of the disc when the viscous stress is proportional to the total gas+radiation+magnetic pressure. Disc parameters are $p_v = p + p_m$, $\alpha_s=0.1$, $\alpha_b=0$, $e=q=0.9$ and $E_0 = 0$. This confirms that a strong magnetic field can stabilise the thermal instability in an eccentric disc if $p_v$ includes the magnetic pressure.} 
\label{magnetic supported}
\end{figure}

\begin{figure}
\centering
\includegraphics[trim=0 0 0 0,clip,width=0.9\linewidth]{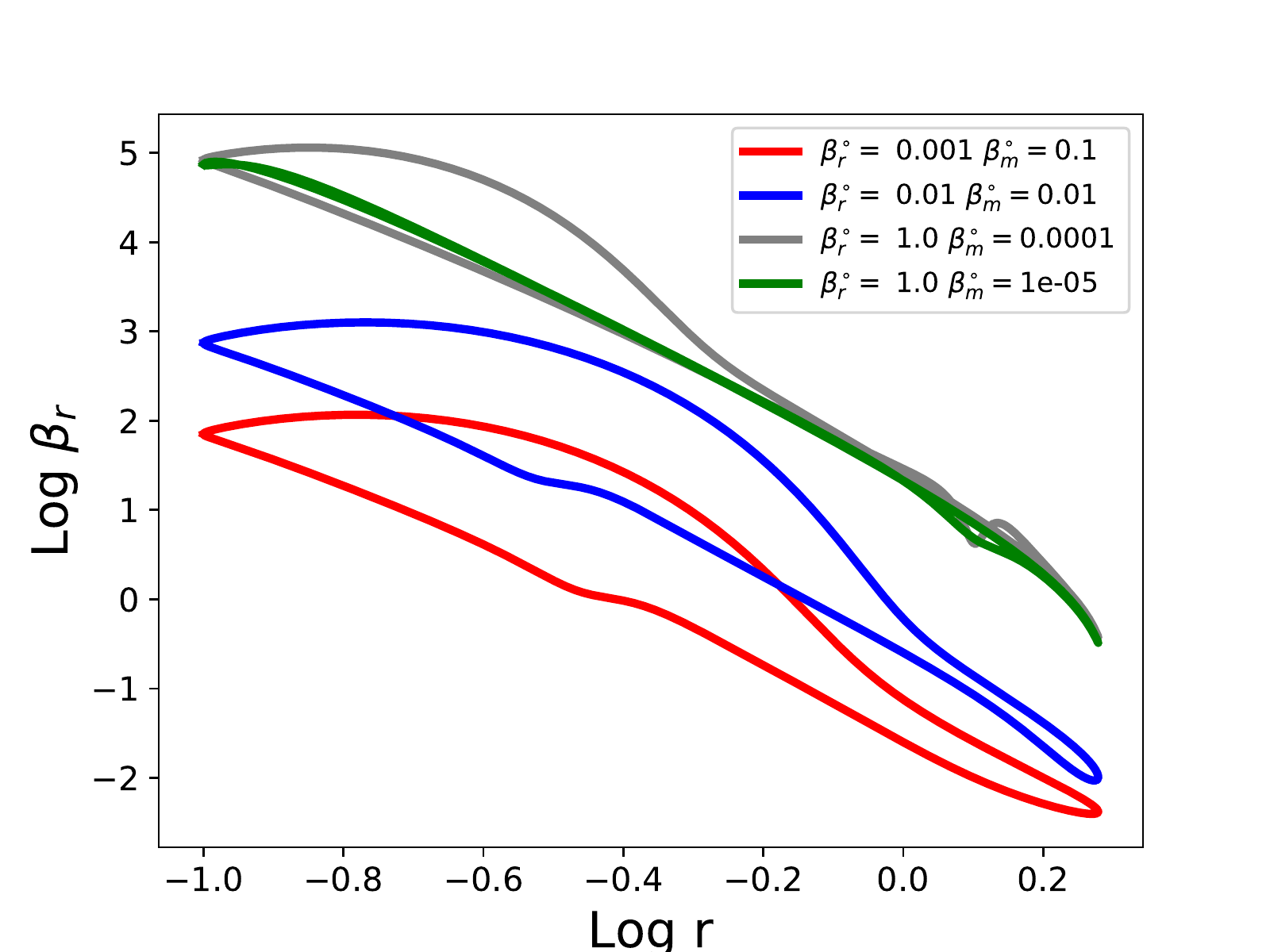}
\caption{Variation of $\beta_r$ around the orbit for each model in Figure \ref{magnetic supported}.} 
\label{magnetic supported betaplot}
\end{figure}

\begin{figure}
\centering
\includegraphics[trim=0 0 0 0,clip,width=0.9\linewidth]{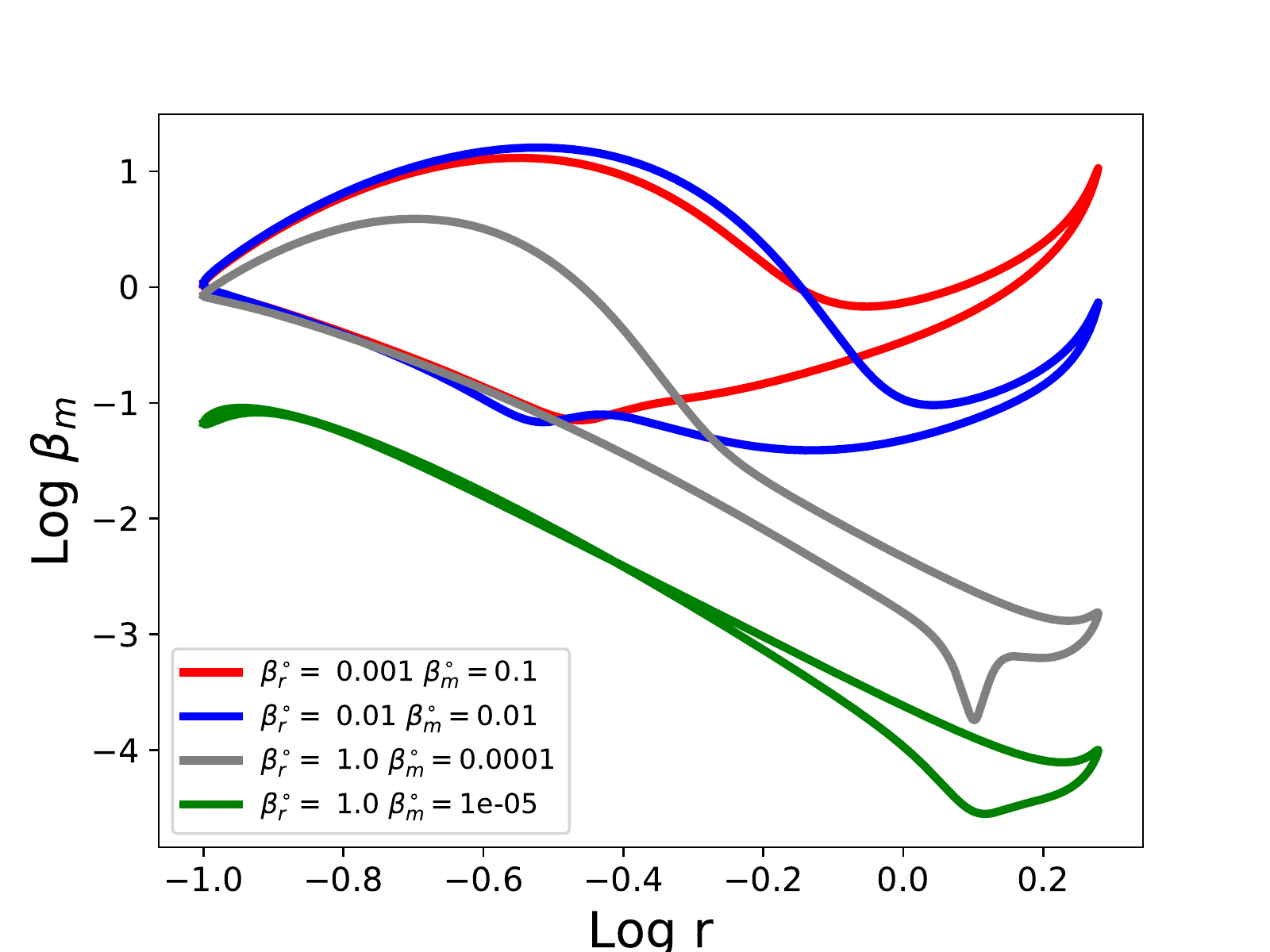}
\caption{Variation of the plasma-$\beta$ around the orbit for each model in Figure \ref{magnetic supported}.} 
\label{magnetic supported plasmabetaplot}
\end{figure}

Figures \ref{magnetic supported}-\ref{magnetic supported plasmabetaplot} show the variation of the scale height, $\beta_r$ and $\beta_m$ for a disc with $p_v = p + p_m$, $\alpha_s=0.1$, $\alpha_b=0$, $e=q=0.9$ and $E_0 = 0$. Here we find that, with a strong enough magnetic field, we can obtain thermally stable solutions despite the dependence of the stress on the radiation pressure. Generally for thermal stability the magnetic field needs to dominate (over radiation pressure) over part of the orbit. Having such a strong horizontal magnetic field over a sizable fraction of the orbit may lead to flux expulsion through magnetic buoyancy, an effect we do not treat here. If the magnetic field is too weak, however, we encounter the thermal instability similar to the hydrodynamic radiation pressure dominated discs when $p_v=p$. 

\begin{figure}
\includegraphics[trim=0 0 0 0,clip,width=\linewidth]{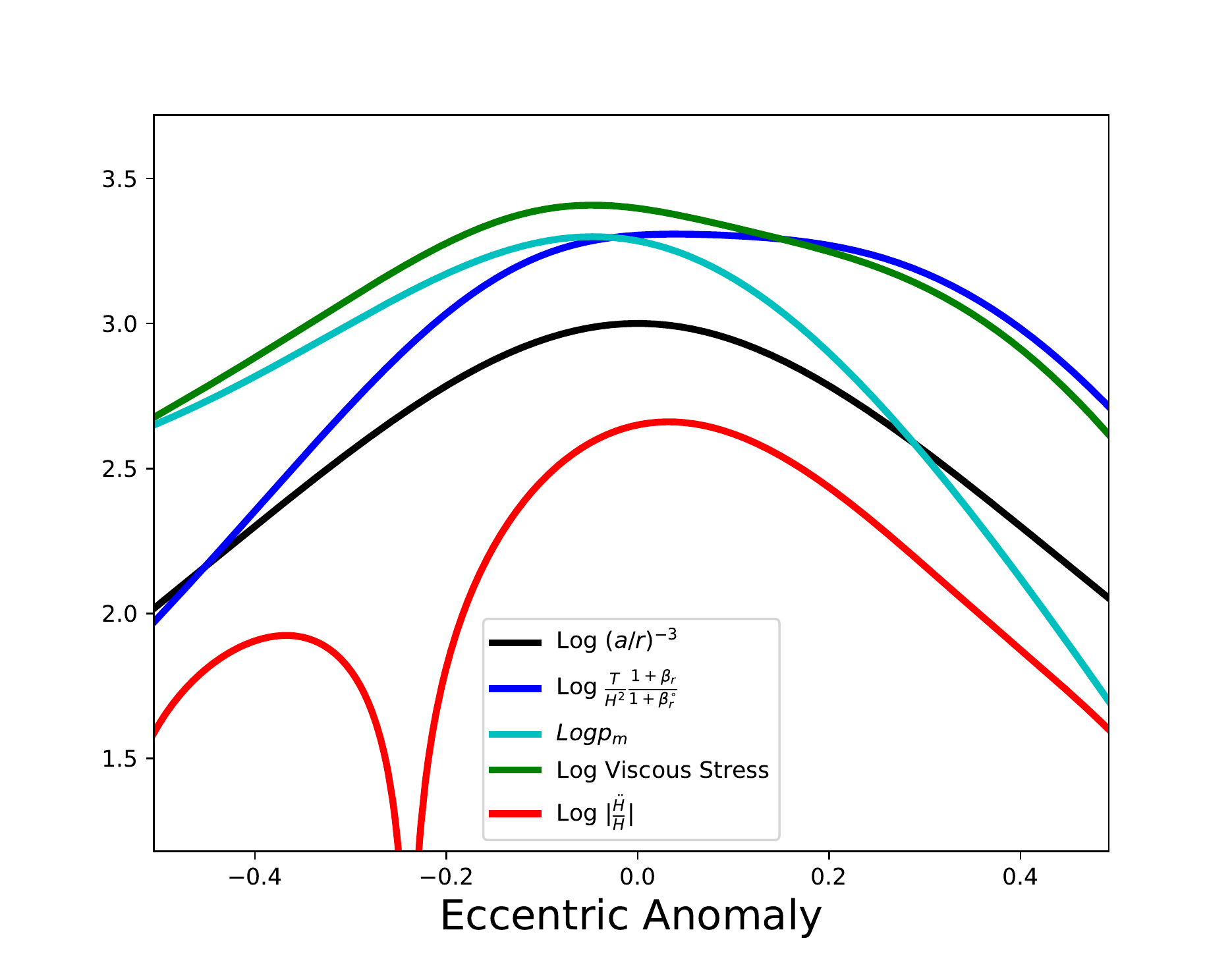}
\caption{Magnitude of terms for a disc with $p_v = p + p_m$, $\beta_r^{\circ} = 10^{-3}$, $\beta_m^{\circ} = 0.1$ $\alpha_s=0.1$, $\alpha_b=0$, $e=q=0.9$ and $E_0 = 0$. The viscous stress exceeds the magnetic, gas and radiation pressures in the nozzle, during pericentre passage.} 
\label{force balance pv pm}
\end{figure}

Part of the motivation for introducing the magnetic field was to regularise some of the extreme behaviour encountered at pericentre. Unfortunately, while the prescriptions $p_v = p_m$ and $p_v = p + p_m$ are promising as a way of taming the thermal instability they exhibit the same extreme behaviour that the hydrodynamic models possess. In particular when $p_v = p_m$ the solutions exhibit extreme compression at pericentre, while for the more magnetised discs (with either $p_v = p_m$ or $p_v = p + p_m$) we again encounter the issue of the viscous stresses being comparable to or exceeding the pressure (including the magnetic pressure). See, for example, Figure \ref{force balance pv pm} which shows that the viscous stresses exceed the magnetic, gas and radiation pressures during pericentre passage.

\section{Nonlinear constitutive model for the magnetic field} \label{constituative model}

The model considered in Section \ref{magnetic pressure effect} has a number of drawbacks. The first is that the viscous stress and the coherent magnetic field are treated as separate physical effects when they are in fact intrinsically linked (although subsection \ref{mag dependant stress} partially addresses this issue). Secondly, the turbulent magnetic field, responsible for the effective viscosity, cannot store energy. Lastly the model neglects resistive effects and, while nonideal MHD effects would be weak if the flow were strictly laminar, the turbulent cascade should always move magnetic energy to scales on which nonideal effects become important. Thus the coherent magnetic field should be affected by some dissipative process.

To address these issues we consider a model of the (modified) Maxwell stress where the magnetic field is forced by a turbulent emf and relaxes to a isotropic field proportional to some pressure $p_v$ on a timescale $\tau$. While the ``turbulence'' in this model acts to isotropise the magnetic field, the presence of the background shear flow feeds off the quasi-radial field component and produces a highly anisotropic field that is predominantly quasi-toroidal. A possible justification for this model based on a stochastically forced induction equation is given in Appendix \ref{mag deriv}. This model has much in common with \citet{Ogilvie03}, but does not solve for the Reynolds stress explicitly.

The Maxwell stress in this model evolves according to

\begin{equation}
\mathcal{D} M^{i j} = -(M^{i j}  - \mathcal{B} p_v g^{i j})/\tau \quad ,
\label{m stress equation}
\end{equation}
where $g^{ij}$ is the metric tensor and $\mathcal{B}$ is a nondimensional parameter controlling the strength of forcing relative to $p_v$. $\mathcal{B}$ can can be taken to be constant by absorbing any variation into the definition of $p_v$. $\mathcal{D}$ is the operator from \citet{Ogilvie01} (a type of weighted Lie derivative) which acts on a rank (2,0) tensor by

\begin{equation}
\mathcal{D} M^{i j} = D M^{i j} - 2 M^{k (i} \nabla_k u^{j)} + 2M^{i j} \nabla_k u^k \quad .
\end{equation}
As noted in \citet{Ogilvie01}, $\mathcal{D} M^{i j} = 0$ is the equation for the evolution of the (modified) Maxwell stress for a magnetic field which satisfies the ideal induction equation; it states that the magnetic stress is frozen into the fluid. 

We adopt the following prescription for the relaxation time:

\begin{equation}
 \tau = \mathrm{De}_0 \frac{1}{\Omega_z} \sqrt{\frac{p_v}{M}} \quad ,
\label{We relatition}
\end{equation}
where $\Omega_z = \sqrt{G M_{\bullet}/r^3}$ is the vertical oscillation frequency and $\mathrm{De}_0$ is a dimensionless constant; this matches the functional form for the relaxation time $\tau$ given in the compressible version of \citet{Ogilvie03}. In subsequent equations it will be useful to express this relaxation time as a Deborah number $\mathrm{De} = n \tau$, a dimensionless number used in viscoelastic fluids that is the ratio of the relaxation time to some characteristic timescale of the flow. When $\tau$ is given by Equation \ref{We relatition} then Equation \ref{m stress equation} corresponds to the equation for the (modified)-Maxwell stress given in \citet{Ogilvie03} if the Reynolds stress is isotropic and proportional to some pressure $p_v$. One emergent property of such a stress model is that the stress will naturally scale with magnetic pressure, as the latter is the trace of the former (see Appendix \ref{stress model prop}).

From this stress model, we have a nondimensional heating rate,

\begin{equation}
f_{\mathcal{H}} = \frac{1}{2 \mathrm{De}} \left( \frac{M}{p_v} - 3 \mathcal{B} \right) \quad,
\end{equation}
which ensures that magnetic energy loss/gained via the relaxation terms in Equation \ref{m stress equation} is converted to/from the thermal energy (this is shown in Appendix \ref{circular disc}). Thus energy is conserved within the disc, although it can be lost radiatively from the disc surface.

In Appendix \ref{circular disc} we obtain the hydrostatic solutions for a circular disc. If $p_v$ is independent of $M$ the vertical equation of motion, rescaled by this reference circular disc, is

\begin{equation}
\frac{\ddot{H}}{H} = -(1 - e \cos E)^{-3} + \frac{T}{H^2} \frac{1 + \beta_r}{1 + \beta_{r}^{\circ} } \frac{ \Biggl( 1 + \frac{1}{2} \frac{M}{p} - \frac{M^{z z}}{p} \Biggr)}{\left[ 1 + \mathcal{B} \frac{P_v^{\circ}}{P^{\circ}} \left(\frac{1}{2} + \frac{9}{4} \mathrm{De}_0^2 \frac{P_v^{\circ}}{P^{\circ}} \right) \right]} ,
\label{equipart momentum}
\end{equation}
while the thermal energy equation is

\begin{align}
\begin{split}
  \dot{T} &= - (\Gamma_3 - 1) T \left(\frac{\dot{J}}{J} + \frac{\dot{H}}{H} \right) \\
  &+ (\Gamma_3 - 1) \frac{1 + \beta_r}{1 + 4 \beta_r} T \left[ \frac{1}{2 \mathrm{De}} \left( \frac{M}{p} - 3 \mathcal{B} \frac{p_v}{p} \right) - \mathcal{C}^{\circ}  \frac{1 + \beta_{r}^{\circ} }{1 + \beta_r} J^2 T^{3} \right] ,
 \end{split}
 \label{equipart thermal}
\end{align}
where we have introduced a reference cooling rate,

\begin{align}
\begin{split}
\mathcal{C}^{\circ} &= \frac{9}{4} \mathcal{B} \mathrm{De}^{\circ} \frac{P^{\circ}_v}{P^{\circ}} \\
&= \left(\frac{3}{2} \right)^{3/2} \mathcal{B}^{1/2} \mathrm{De}_0 \left( 1 + \sqrt{1 + 2 \frac{\mathrm{De}_0^2}{\mathcal{B}}} \right)^{-1/2} \frac{P^{\circ}_v}{P^{\circ}} \quad .
 \end{split}
\end{align}
Here $\mathrm{De}^{\circ}$ is the equilibrium Deborah number in the reference circular disc, which is in general different from $\mathrm{De}_0$.

We solve these equations along with the equations for the evolution of the stress components,

\begin{align}
    \dot{M}^{\lambda \lambda} &+ 2 \left(\frac{\dot{J}}{J} + \frac{\dot{H}}{H} \right) M^{\lambda \lambda} = -(M^{\lambda \lambda}  -  \mathcal{B} p_v g^{\lambda \lambda})/\mathrm{De} , \\
    \dot{M}^{\lambda \phi} & - M^{\lambda \lambda} \Omega_{\lambda} - M^{\lambda \phi} \Omega_{\phi}  + 2 \left(\frac{\dot{J}}{J} + \frac{\dot{H}}{H} \right) M^{\lambda \phi} \nonumber  \\
   &= -(M^{\lambda \phi}  -  \mathcal{B} p_v g^{\lambda \phi})/\mathrm{De} , \\
    \dot{M}^{\phi \phi} &- 2 M^{\lambda \phi} \Omega_{\lambda} - 2 M^{\phi \phi} \Omega_{\phi}  + 2 \left(\frac{\dot{J}}{J} + \frac{\dot{H}}{H} \right) M^{\phi \phi} \nonumber  \\
    &= -(M^{\phi \phi}  -  \mathcal{B} p_v g^{\phi \phi})/\mathrm{De} , \\
    \dot{M}^{z z} &+ 2 \frac{\dot{J}}{J} M^{z z} = -(M^{z z}  -  \mathcal{B} p_v g^{z z})/\mathrm{De}  \quad.
\end{align}
We solve for these stress components in the $(\lambda,\phi)$ coordinate system of \citet{Ogilvie01} as this simplifies the metric tensor. We can do this as, apart from $M^{zz}$ (which is the same in both coordinate systems), our equations only depend on $M^{i j}$ through scalar quantities.

\begin{figure}
\centering
\includegraphics[trim=0 0 0 0,clip,width=0.9\linewidth]{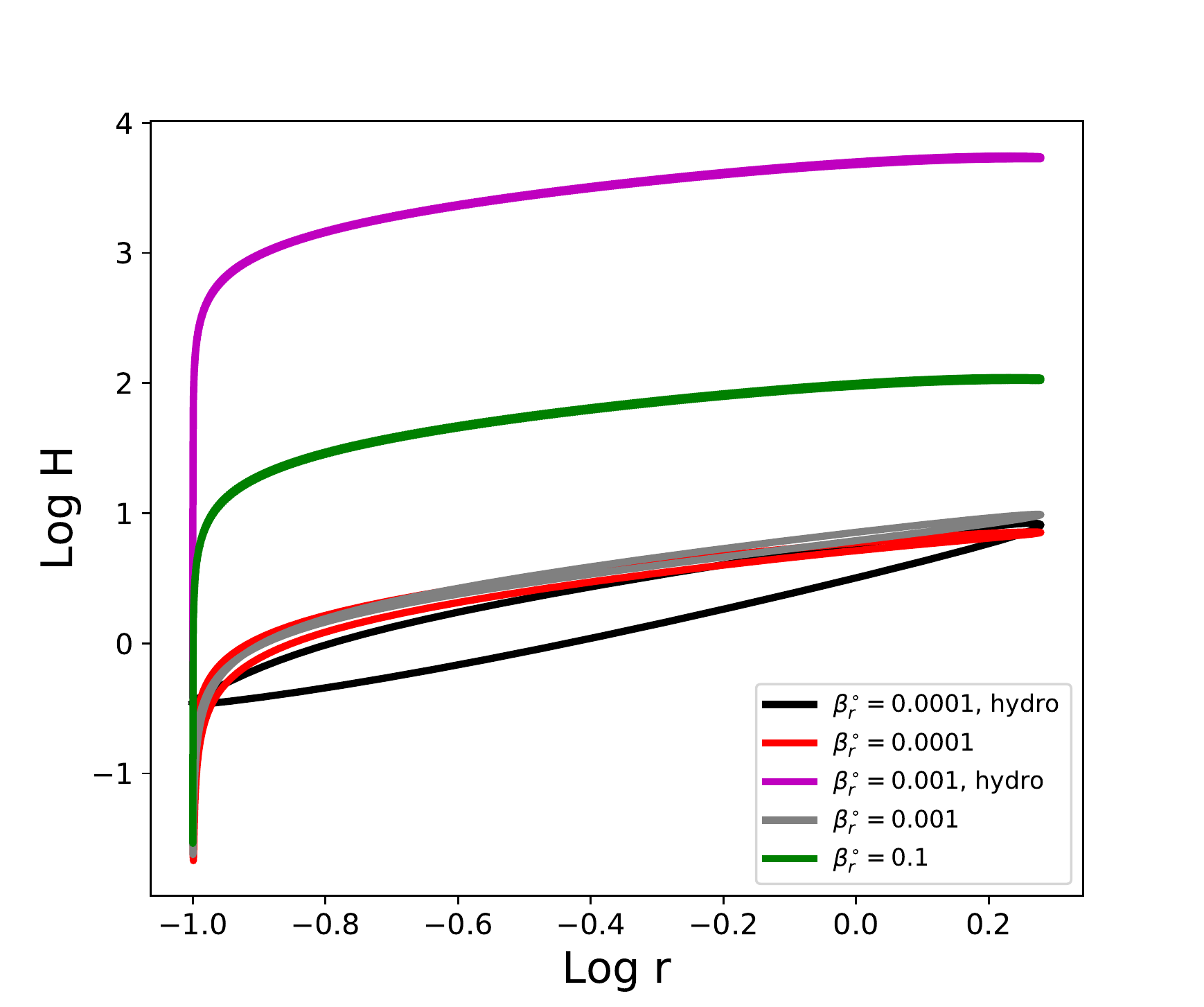}
\caption{Variation of the scale height of the disc with radiation + gas pressure with different $\beta_r^{\circ}$ using our modified Maxwell stress prescription. Disc parameters are $e=q=0.9$ and $E_0 = 0$ and $p_v = p_g$; $\alpha$-discs have $\alpha_s=0.1$, $\alpha_b = 0$ while the Maxwell stress prescription has $\mathrm{De}=0.5$, $\mathcal{B} = 0.1$. Black line is an $\alpha$-disc with $\beta_{r}^{\circ} = 10^{-4}$; red line has $\beta_{r} = 10^{-4}$ with the  Maxwell stress prescription, magenta line is an $\alpha$-disc with $\beta_{t}^{\circ} = 10^{-3}$, grey line has $\beta_{r} = 10^{-3}$ with the Maxwell stress prescription, green line has $\beta_{r} = 10^{-3}$ with the  Maxwell stress prescription.} 
\label{equipartion highbeta comparison}
\end{figure}

\begin{figure}
\centering
\includegraphics[trim=0 0 0 0,clip,width=0.9\linewidth]{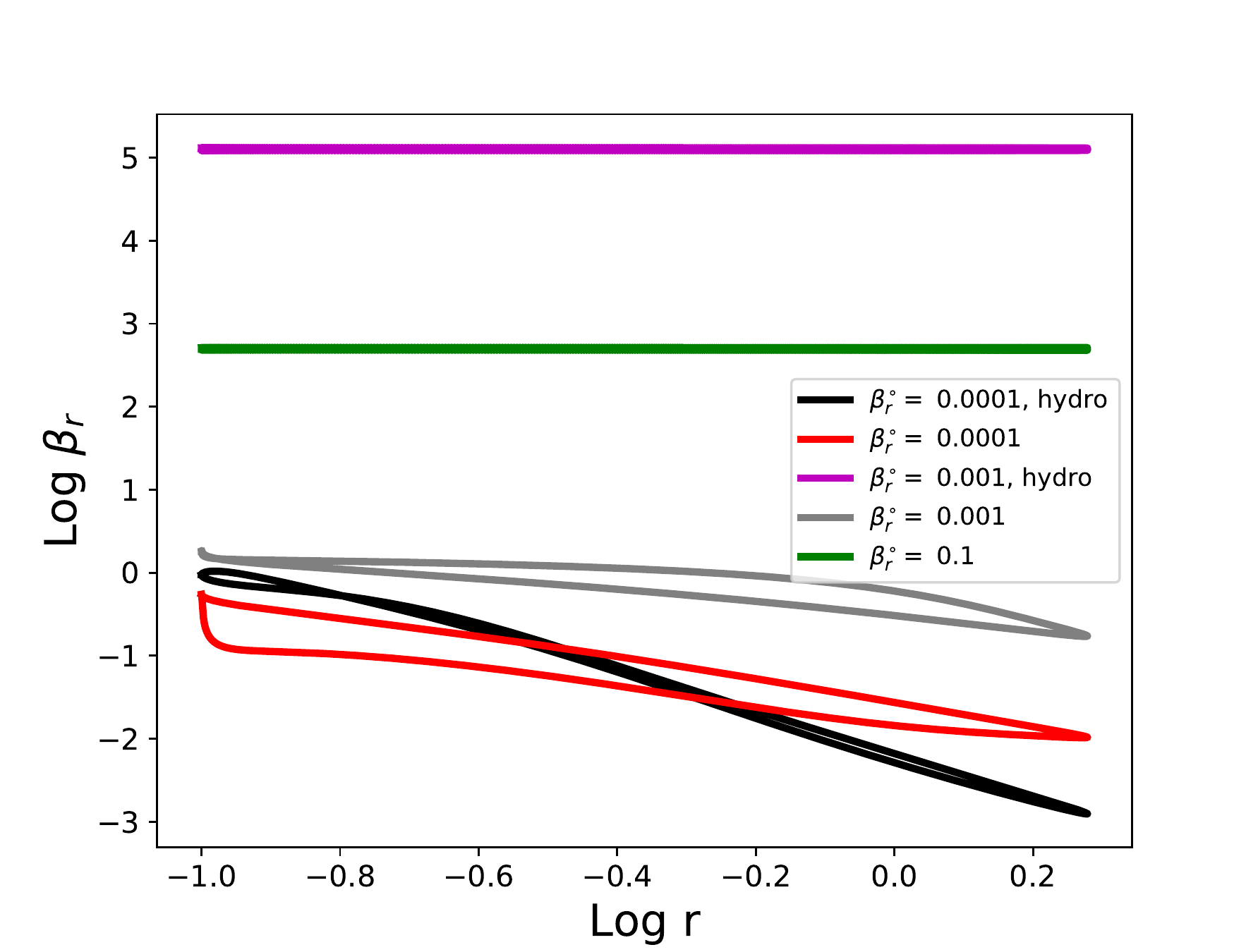}
\caption{Variation of $\beta_r$ around the orbit for each model in Figure \ref{equipartion highbeta comparison}.} 
\label{equipartion highbeta comparison betaplot}
\end{figure}

\begin{figure}
\centering
\includegraphics[trim=0 0 0 0,clip,width=0.9\linewidth]{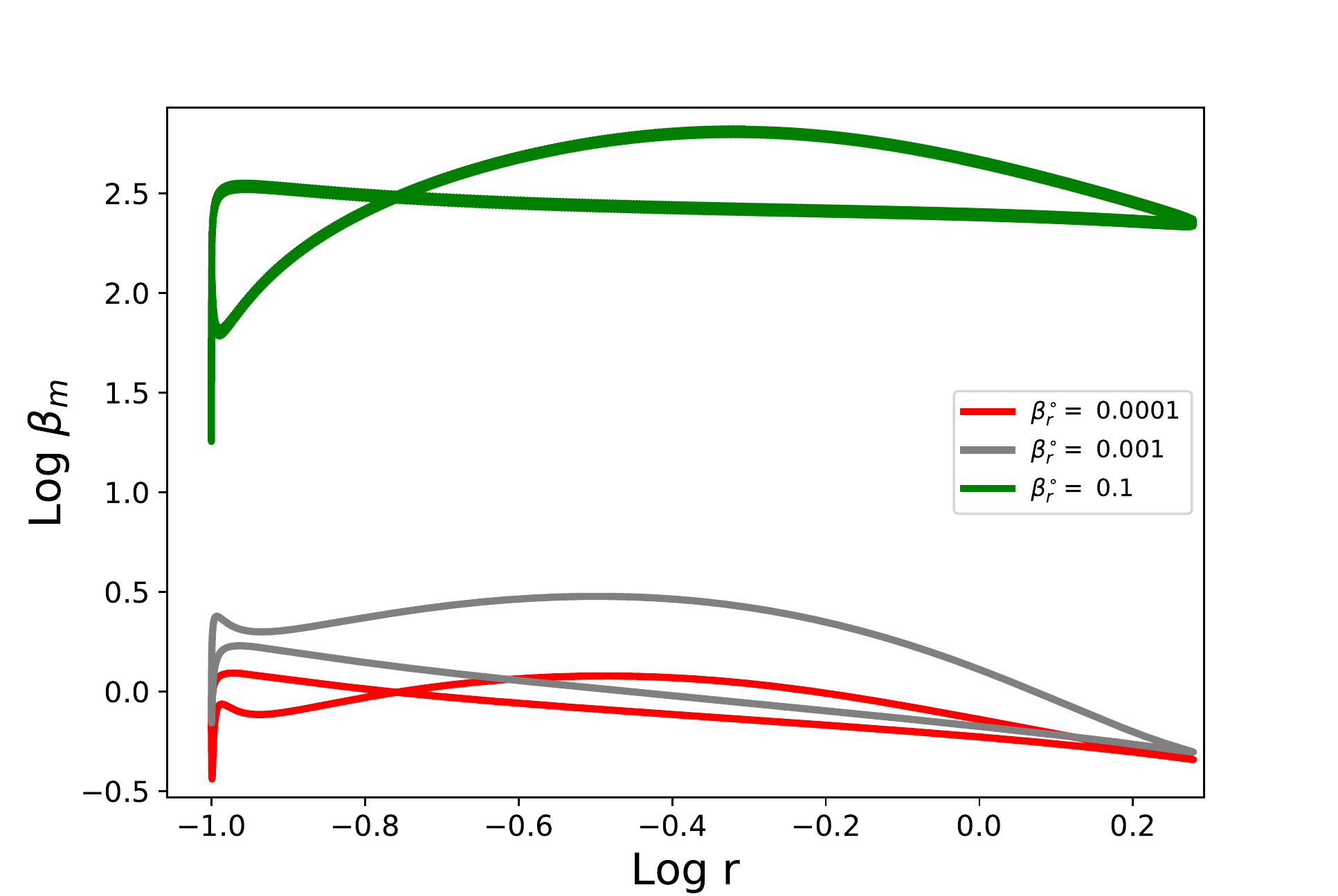}
\caption{Variation of the plasma-$\beta$ around the orbit for each model in Figure \ref{equipartion highbeta comparison}.} 
\label{equipartion highbeta comparison plasmabetaplot}
\end{figure}

Figures \ref{equipartion highbeta comparison}-\ref{equipartion highbeta comparison plasmabetaplot} show the variations of the scale height, $\beta_r$ and plasma beta (defined as $\beta_m = \frac{2 p}{M}$) around the orbit for a disc with $p_v = p_g$, $\mathrm{De}_0=0.5$, $\mathcal{B} = 0.1$, $e=q=0.9$ and $E_0 = 0$. Like the ideal induction equation model of Section \ref{breathinmode}, the coherent magnetic field has a stabilising effect on the dynamics. The effect is not as strong as that seen in the ideal induction equation model as, in that model, we could choose $\beta_{m}^{\circ}$ so as to achieve a much stronger field than achieved by the constitutive model here. Compared with the ideal induction equation model the plasma-$\beta$ is more uniform around the orbit; there is still an abrupt decrease in the plasma-$\beta$ near pericentre, which highlights the importance of the magnetic field during pericentre passage.

\begin{figure}
\includegraphics[trim=0 0 0 0,clip,width=\linewidth]{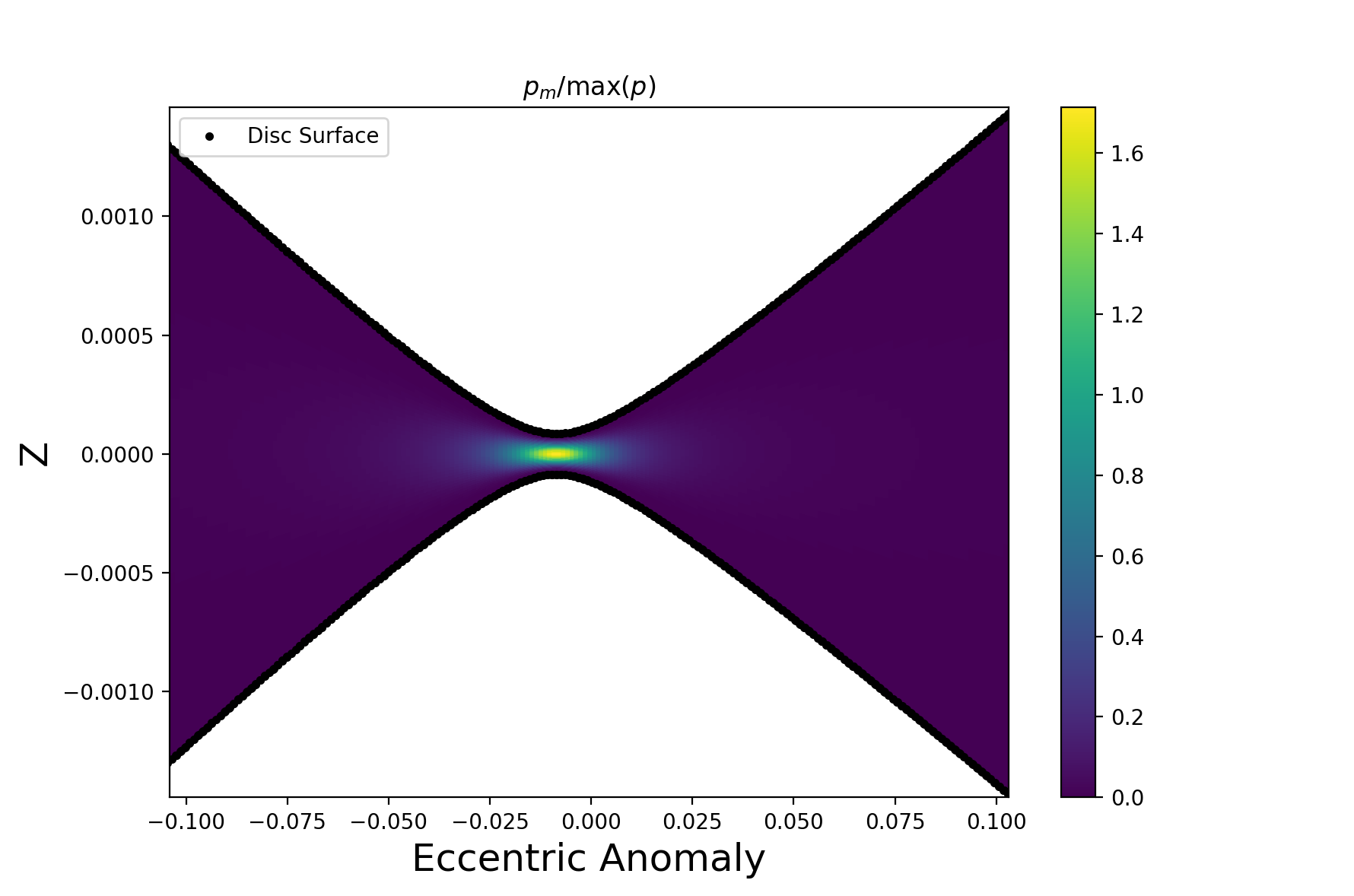}
\caption{Pericentre passage for a disc using our modified Maxwell stress prescription, with $p_v = p_g$, $\beta_{r}^{\circ} = 10^{-3}$, $\mathrm{De}_0 = 0.5$, $\mathcal{B} = 0.1$, $e=q=0.9$ and $E_0 = 0$ showing the magnetic pressure ($M/2$). As with the ideal induction equation model the magnetic field is highly concentrated in the nozzle. The nozzle is very nearly symmetric, although it is slightly offset from pericentre, indicating weak dissipation.}  
\label{equi nozzle}
\end{figure}

Figure \ref{equi nozzle} shows the pericentre passage for a disc with $p_v = p_g$, $\beta_{r}^{\circ} = 10^{-3}$, $\mathrm{De}_0 = 0.5$, $\mathcal{B} = 0.1$, $e=q=0.9$ and $E_0 = 0$. As with the ideal induction equation model, the magnetic pressure is extremely concentrated within the nozzle and near to the midplane. The nozzle is far more symmetric compared to the ideal induction equation model as the weaker field means that the disc is in a modified form of the nearly adiabatic radiation pressure dominated state. 

In addition to considering the situation where the fluctuation pressure scales with the gas pressure $p_v = p_g$, we also considered $p_v=p+p_m$. As in the ideal induction equation model we found it is possible to stabilise the thermal instability with a strong enough magnetic field; however we found that this requires fine tuning of $\mathrm{De}_0$ and $\mathcal{B}$, for which there is no obvious justification. However, instead of stabilising the thermal instability, it is possible to delay its onset by choosing a small enough $\mathrm{De}_0$, so that the thermal runaway occurs on a timescale much longer than the orbital time (occurring after $\sim 1000$ orbits). However, these solutions never settle down into a periodic (or nearly periodic) solution and instead have a long phase of quasi-periodic evolution, where the mean scale height remains close to its initial value, before eventually experiencing thermal runaway. A quasi-periodic solution of our model is not self consistent, so the possibility that the thermal instability is delayed in the nonlinear-constitutive MRI model needs to be explored using an alternative method. 

The possibility that the thermal instability stalls or is delayed has some support from simulations looking at the thermal stability of MRI active discs \citep{Jiang13,Ross17}. In both these papers it was found that the disc was quasi-stable, with thermal instability occurring when a particularly large turbulent fluctuation caused a strong enough perturbation away from the equilibrium. The (modified) Maxwell stress considered here is equal to the expectation value of a modified Maxwell stress which is stochastically forced by fluctuation with amplitude proportional to $p_v$ (see Equation \ref{stocastic maxwell stress} of Appendix \ref{mag deriv} and discussion therein), so it is possible that our stress model captures the thermal quasi-stability seen in \citet{Jiang13} and \citet{Ross17} in some averaged sense. The possibility that the thermal instability is delayed or slowed is particularly relevant for TDEs which are inherently transient phenomena -- if the timescale for thermal runaway is made long enough then eccentric TDE discs maybe thermally stable over the lifetime of the disc.

\section{Discussion}

\subsection{Stability of the solutions} \label{stabil}

As discussed in Paper I, one advantage of our solution method is that the solutions it finds are typically nonlinear attractors (or at least long lived transients) and so are stable against (nonlinear) perturbations to the solution variables ($H$,$\dot{H}$, $T$ and $M^{i j}$ when present). Generally we expect such perturbations to damp on the thermal time or faster. Instabilities such as the thermal instability manifest as a  failure to converge to a $2 \pi-$periodic solution.

For the ideal induction equation model, our method cannot tell us about the stability of the solution to perturbations to the horizontal magnetic field. Showing this would require a separate linear stability analysis. Perturbations to the vertical field typically have no influence on the dynamics of the disc vertical structure.

However, for the constitutive model, perturbations to the magnetic field are encapsulated in perturbations to $M^{i j}$ so these solutions are stable against (large scale) perturbations to the magnetic field. This is most likely because, unlike the ideal induction equation, dissipation acts on the magnetic field. 

Our solution method doesn't tell us about the stability of our solutions to short wavelength (comparable or less than the scale height) perturbations to our system. So our disc structure could be unstable to such perturbations. Like the hydrodynamic solutions in Paper I, it is likely our discs are unstable to  the parametric instability  \citep{Papaloizou05a,Papaloizou05b,Wienkers18,Barker14,Pierens20}. Additionally if, as assumed, turbulence in highly eccentric discs is caused by the MRI then there must be perturbations to the magnetic field in the ideal induction equation model which are unstable.

Interestingly the simulations of \citet{Sadowski16} found that the strength of turbulence in magnetised and unmagnetised TDE discs is broadly comparable, something that would not be expected in a circular disc. \citet{Sadowski16} suggested that a hydrodynamic instability might be responsible for the disc turbulence. The discs considered by \citet{Sadowski16} still have appreciable eccentricity at the end of their simulation (with $e \approx 0.2$) so an obvious contender would be the parametric instability feeding off the disc eccentricity and breathing mode. 


\subsection{Resistivity and dynamo action} \label{dynamo}

Even when the magnetic field in our models does a good job of resisting the collapse of the disc, the stream will still be highly compressed at pericentre. The highly compressed flow combined with a very strong field (with $\beta_m \ll 1$) makes the nozzle a prime site for magnetic reconnection. This will require that magnetic field lines on neighbouring orbits can have opposite polarities. Our solutions are agnostic to the magnetic field polarity and, in principle, support this possibility.

The simulations of \citet{Guillochon17} suggest that the initial magnetic field in the disc will be (quasi-)toroidal with periodic reversals in direction. When such a field is compressed both horizontally and vertically during pericentre passage, neighbouring toroidal magnetic field lines of opposite polarity can undergo reconnection, generating a quasi-poloidal field. We thus have a basis for an eccentric $\alpha-\Omega$ dynamo, where strong reconnection in the nozzle creates quasi-poloidal field, which in turn creates a source term in the quasi-toroidal induction equation from the shearing of this quasi-radial field (see Appendix \ref{induction equation}). In highly eccentric TDE discs, with their large vertical and horizontal compression, this dynamo could potentially be quite strong. 

The constitutive model we considered in Section \ref{constituative model} implicitly possesses a dynamo through the source term proportional to $p_v$. This is a small scale dynamo which arises from the action of the turbulent velocity field and will be limited by the kinetic energy in the turbulence. A dynamo closed by reconnection in the nozzle would be a large scale dynamo and the coherent field produced could potentially greatly exceed equipartition with the turbulent velocity field.

Lastly reconnection sites can accelerate charged particles. We would therefore expect that TDEs in which there is a line of sight to the pericentre should include a flux of high-energy particles. From most look angles the line of sight to the pericentre ``bright point'' will be blocked (see also \citet{Zanazzi20}). This will both block the X-ray flux from the disc and any particles accelerated by reconnection in the nozzle. Hence these high energy particles should only be seen in X-ray bright TDEs.

\section{Conclusion} \label{conc}

In this paper we considered alternative models of the Maxwell stress from the standard $\alpha-$prescription applied to highly eccentric TDE discs. In particular we focus on the effects of the coherent magnetic field on the dynamics. We consider two separate  stress models: an $\alpha-$disc with an additional coherent magnetic field obeying the ideal induction equation and a nonlinear constitutive (viscoelastic) model of the magnetic field. In summery our results are

\begin{enumerate}
\item The coherent magnetic field in both models has a stabilising effect on the dynamics, making the gas pressure dominated branch stable at larger radiation pressures and reducing or removing the extreme variation in scale height around the orbit for radiation pressure dominated solutions.
\item The coherent magnetic field is capable of reversing the collapse at pericentre without the presumably unphysically strong viscous stresses seen in some of the hydrodynamic models.
\item For the radiation pressure dominated ideal induction equation model with a moderate magnetic field the dynamics of the scale height is set by the magnetic field (along with gravity and vertical motion) and doesn't feel the effects of gas or radiation pressure. This is because magnetic pressure dominates during pericentre passage which is the only part of the orbit where pressure is important.
\end{enumerate}

At present the behaviour of magnetic fields in eccentric discs is an understudied area. Our investigation suggests that magnetic fields can play an important role in TDE discs and that further work in this area is needed. 

\section*{Acknowledgements}

We thank J. J. Zanazzi, L. E. Held and H. N. Latter for helpful discussions. E. Lynch would like to thank the Science and Technologies Facilities Council (STFC) for funding this work through a STFC studentship. This research was supported by STFC through the grant ST/P000673/1.

\section{Data availability}

The data underlying this article will be shared on reasonable request to the corresponding author.




\bibliographystyle{mnras}
\bibliography{magnetic_eccentric_tde} 




\appendix

\onecolumn

\section{Properties of our Stress model} \label{stress model prop} 
 
In this appendix we show some important properties of our constitutive model for the Maxwell stress.  Restated here for clarity, our model for the (modified) Maxwell stress is
 
\begin{equation}
M^{i j} + \tau \mathcal{D} M^{i j} = \mathcal{B} p_v g^{i j} \quad ,
\label{equi stress consituative appendix}
\end{equation}
where $p_v$ is some pressure which controls the magnitude of the magnetic fluctuations. The Deborah number is given by

\begin{equation}
\mathrm{De} = \tau n =  \mathrm{De}_0 \frac{n}{\Omega_z} \sqrt{\frac{p_v}{M}},
\label{weissberg num}
\end{equation}
where $\Omega_z = \sqrt{G M_{\bullet}/r^3}$ and $\mathrm{De}_0$ is a dimensionless constant. This matches the decay term in the compressible version of \citet{Ogilvie03}. 

\subsection{Viscoelastic behaviour}
 
This model is part of a large class of possible viscoelastic models for $M^{i j}$. The elastic limit $\tau \rightarrow \infty$ of this equation is fairly obvious, corresponding to a magnetic field which obeys the ideal induction equation through the ``freezing in'' of $M^{i j}$ ($\mathcal{D} M^{i j} = 0$). However, to obtain the viscous behaviour responsible for the effective viscosity of circular accretion discs requires more work. The viscous limit is obtained when $\mathrm{De} \ll 1$. For simplicity, in what follows, we shall assume $\tau$ and $p_v$ are independent of the magnetic field ($M^{i j}$).

We propose a series expansion in $\tau$,

\begin{equation}
M^{i j} = \sum_{k = 0}^{\infty} \tau^k M_k^{i j} \quad.
\end{equation} 
This expansion is only likely to be valid in the short $\tau$ limit and may break down for material variations on timescales shorter than $\tau$. With that caveat we can find a series solution for equation \ref{equi stress consituative appendix},

\begin{equation}
M^{i j} =  \mathcal{B} \sum_{k = 0}^{\infty} (- \tau \mathcal{D})^k p_v g^{i j} \quad .
\end{equation}
Keeping the lowest order terms in the expansion we have

\begin{equation}
M^{i j} = \mathcal{B} p_v g^{i j} - \mathcal{B} \tau \mathcal{D} (p_v g^{i j}) + O(\tau^2) \quad .
\end{equation}
The lowest order term is an isotropic stress and evidently a form of magnetic pressure. The operator $\mathcal{D} = D$ when acting on a scalar and when acting on the metric tensor $\mathcal{D} g^{i j} = -2 S^{i j} + 2 \nabla_k u^k g^{i j}$. Using the product rule,

\begin{align}
\begin{split}
M^{i j} &\approx \mathcal{B} p_v g^{i j} + 2 \mathcal{B} \tau p_v S^{i j} - 2 \mathcal{B} \tau ( p_v \nabla_k u^k + D p_v ) g^{i j}  \\
&= \mathcal{B} p_v g^{i j} + 2 \mathcal{B} \tau p_v S^{i j} + 2 \mathcal{B} \tau \left( \left(\frac{\partial p_v}{\partial \ln \rho} \right)_{s} - p_v \right) \nabla_k u^k g^{i j}  - 2 \mathcal{B} \tau \left(\frac{\partial p_v}{\partial s} \right)_{\rho}  (D s) \, g^{i j} \quad.
\end{split}
\label{lin tau expansion}
\end{align}
So the first $O(1)$ term gives rise to a magnetic pressure, the first $O(\tau)$ term is a shear viscous stress, the second is a bulk viscous stress and the final term is an additional nonadiabatic correction which has no obvious analogue in the standard viscous or magnetic models. Higher order terms contain time derivatives of the pressure $p_v$ and the velocity gradients $\nabla_i u^{j}$. This dependence produces a memory effect in the fluid causing the dynamics of the fluid to depend on its history. The fluid has a finite memory and becomes insensitive to (``forgets about") its state at times $\gtrsim \tau$ in the past. 

If $\tau$ and $p_v$ depend on $M^{i j}$ then the terms in equation \ref{lin tau expansion} are modified. However the equation still decomposes into an isotropic magnetic pressure term, a shear stress term $\propto S^{i j}$, a bulk stress term $\propto \nabla_k u^k g^{i j}$ and a non-adiabatic term $\propto (D s)\, g^{i j}$ (as in equation \ref{lin tau expansion}). 

\subsection{Realisability}

In addition to its behaviour in the viscous and elastic limits another necessary property of a model of a Maxwell stress is its realisability from actual magnetic fields. As $M^{i j} = \frac{B^i B^j}{\mu_0}$, $M^{i j}$ must be positive semi-definite. Thus for all positive semi-definite initial conditions $M^{i j} (0)$ our constitutive model equation \ref{equi stress consituative appendix} must conserve the positive semi-definite character of $M^{i j}$. This is equivalent to requiring that the quadratic form $Q = M^{i j} Y_i Y_j$ satisfy $Q \ge 0$ for all vectors $Y_i$, at all points in the fluid.

We will show by contradiction that an initially positive semi-definite $M^{ij}$ cannot evolve into one that is not positive semi-definite. Suppose, to the contrary, that at some point in the flow $Q < 0$ for some vector $X_i$ at some time after the initial state. Then let us consider a smooth, evolving vector field $Y_i$ that matches the vector $X_i$ at the given point and time. The corresponding quadratic form $Q$ is then a scalar field that evolves according to

\begin{align}
\begin{split}
\mathcal{D} Q &= Y_i Y_j \mathcal{D} M^{i j} + M^{i j} \mathcal{D} (Y_i Y_j) \\
&= \left(\mathcal{B} p_v Y^2 - Q \right)/\tau + M^{i j} \mathcal{D} Y_i Y_j \quad . \\
\end{split}
\end{align}

By assumption, $Q$ is initially positive and evolves continuously to a negative value at the given later time. Therefore $Q$ must pass through zero at some intermediate time, which we denote by $t = 0$ without loss of generality. We can also assume, without loss of generality, that the vector field evolves according to $\mathcal{D}Y_i = 0$, which means that it is advected by the flow. The equation for $Q$ then becomes

\begin{equation}
D Q = \left(\mathcal{B} p_v Y^2 - Q \right)/\tau \quad,
\end{equation}
where we have made use of the fact that $\mathcal{D} = D$ when acting on a scalar. Within the disc we expect $p_v > 0$, additionally as $M^{i j}$ is positive semi-definite $M \ge 0$. Thus at $t=0$, $Q = 0$ and the time derivative of $Q$ is given by

\begin{equation}
D Q |_{t = 0} = \mathcal{B} p_v Y^2/\tau  \ge 0 \quad .
\end{equation}
This contradicts the assumption that $Q$ passes through zero from positive to negative at $t=0$. We conclude that $M^{ij}$ remains positive semi-definite if it is initially so.

\subsection{Energy Conservation} 

In order that the interior of our disc conserve total energy we need to derive the appropriate magnetic heating/cooling rate so that energy lost/gained by the magnetic field is transferred to/from the thermal energy. The MHD total energy equation with radiative flux is

\begin{equation}
\partial_t \left[ \rho \left( \frac{\mathbf{u}^2}{2} + \Phi + e \right) + \frac{\mathbf{B}^2}{2 \mu_0} \right] + \nabla \cdot \left[\rho \mathbf{u} \left( \frac{\mathbf{u}^2}{2} + \Phi + h \right) + \mathbf{u} \frac{\mathbf{B}^2}{\mu_0} - \frac{1}{\mu_0} (\mathbf{u} \cdot \mathbf{B}) \mathbf{B} + \mathbf{F} \right] = 0 \quad ,
\end{equation}
where $e$ is the specific internal energy and $h=e+p/\rho$ is the specific enthalpy. In terms of the modified Maxwell stress,

\begin{equation}
\partial_t \left[ \rho \left( \frac{u^2}{2} + \Phi + e \right) + \frac{M}{2} \right] + \nabla_i \left[\rho u^i \left( \frac{u^2}{2} + \Phi + h \right) + u^i M  - u_j M^{i j} + F^i \right] = 0 \quad .
\end{equation}
From this we deduce the thermal energy equation,

\begin{equation}
\rho T D s = M^{i j} S_{i j} - \frac{1}{2} (D M + 2 M \nabla_i u^i) - \nabla_i F^i \quad .
\label{entrop eq}
\end{equation}
Using the constitutive relation, Equation \ref{equi stress consituative appendix}, we obtain

\begin{equation}
\rho T D s = \frac{1}{2 \tau} \left ( M - 3 \mathcal{B} p_v \right) - \nabla_i F^i \quad ,
\end{equation}
so we have the nondimensional heating rate,

\begin{equation}
f_{\mathcal{H}} = \frac{1}{2 \mathrm{De}} \left( \frac{M}{p_v} - 3 \mathcal{B} \right) \quad .
\end{equation}

Substituting Equation \ref{lin tau expansion} into Equation \ref{entrop eq} we recover, in the viscous limit, terms proportional to $S^{i j} S_{i j}$ and $(\nabla_i u^i)^2$ which act like a viscous heating rate.

\section{Stress model behaviour in a circular disc} \label{circular disc}

In this appendix we consider the behaviour of our nonlinear constitutive model in a circular disc. We derive the reference circular disc, with respect to which our models are scaled. For a circular disc, the fixed point of equation \ref{m stress equation} is

\begin{align}
M^{R R} &= M^{z z} = \mathcal{B} p_v \\
R M^{R \phi} &= -\frac{3}{2} \mathrm{De} \mathcal{B} p_v \\
R^2 M^{\phi \phi} &= \left(1 + \frac{9}{2} \mathrm{De}^2 \right) \mathcal{B} p_v,
\end{align}
which results in a magnetic pressure of
\begin{equation}
p_m = \frac{1}{2} M = \left(\frac{3}{2} + \frac{9}{4} \mathrm{De}^2 \right) \mathcal{B} p_v \quad .
\label{M equation}
\end{equation}
As $\mathrm{De}$ and $p_v$ can depend on $M$ this equation needs to be solved to determine the equilibrium $M$. When $p_v$ is independent of $M$ this yields

\begin{equation}
M = \frac{3}{2} \left( 1 + \sqrt{1 + 2 \frac{\mathrm{De}_0^2}{\mathcal{B}}} \right) \mathcal{B} p_v  \quad,
\label{circular magnetic pressure}
\end{equation}
while for $p_v = p + p_m$ we obtain a quadratic equation for $p_m$,

\begin{equation}
    \left[ 2 - \mathcal{B} \left(3 + \frac{9}{4} \mathrm{De}_0^2 \right) \right] p_m^2 = \mathcal{B} p \left(3 + \frac{9}{2} \mathrm{De}_0^2\right) p_m + \frac{9}{4} \mathrm{De}_0^2 \mathcal{B} p^2 .
\end{equation}
For physical solutions ($p,p_m > 0$) the right hand side is positive. Therefore we require $2 > \mathcal{B} \left(3 + \frac{9}{4} \mathrm{De}_0^2 \right)$ in order that $p_m$ is real. This equation has a singular point at $2 = \mathcal{B} \left(3 + \frac{9}{4} \mathrm{De}_0^2 \right)$, which results in a negative magnetic pressure. Solving for the magnetic pressure,

\begin{equation}
p_m = \frac{3}{2} \mathcal{B} p \frac{1 + \frac{3}{2} \mathrm{De}_0^2 \pm \sqrt{1 + 2 \frac{\mathrm{De}_0^2}{\mathcal{B}}}}{2 - \mathcal{B} \left(3 + \frac{9}{4} \mathrm{De}_0^2 \right)} \quad.
\end{equation}
Given the requirement that $2 > \mathcal{B} \left(3 + \frac{9}{4} \mathrm{De}_0^2 \right)$, the negative root always results in a negative magnetic pressure and is thus unphysical.

In addition to the equilibrium values of $M^{i j}$, the circular reference disc obeys hydrostatic and thermal balance. The equation for hydrostatic equilibrium in a circular disc is

\begin{equation}
\frac{P}{\Sigma H^2} \left(1 + \frac{M}{2 p} - \frac{M_{z z}}{p} \right) = n^2 \quad ,
\label{hydrostatic}
\end{equation}
while the equation for thermal balance is

\begin{equation}
\mathcal{C}^{\circ} = f_{\mathcal{H}} \frac{P^{\circ}_v}{P^{\circ}} = \frac{9}{4} \mathcal{B} \mathrm{De}^{\circ} \frac{P^{\circ}_v}{P^{\circ}} \quad,
\label{thermal balance}
\end{equation}
where

\begin{equation}
\mathrm{De}^{\circ} = \mathrm{De}_0 \sqrt{\frac{p_v^{\circ}}{M^{\circ}}} \quad .
\end{equation}

Taking the solution to equations \ref{hydrostatic}-\ref{thermal balance} which has $\beta_r = \beta_r^{\circ}$ and substituting this into equations \ref{scale height equation} and \ref{thermal energy equation T} as the circular reference state we obtain

\begin{equation}
\frac{\ddot{H}}{H} = -(1 - e \cos E)^{-3} + \frac{T}{H^2} \frac{1 + \beta_r}{1 + \beta_{r}^{\circ} } \frac{ \Biggl( 1 + \frac{1}{2} \frac{M}{p} - \frac{M^{z z}}{p} \Biggr)}{\left[ 1 + \mathcal{B} \frac{P_v^{\circ}}{P^{\circ}} \left(\frac{1}{2} + \frac{9}{4} (\mathrm{De}^{\circ})^2 \right) \right]} \quad ,
\end{equation}

\begin{equation}
 \dot{T} = - (\Gamma_3 - 1) T \left(\frac{\dot{J}}{J} + \frac{\dot{H}}{H} \right) + (\Gamma_3 - 1) \frac{1 + \beta_r}{1 + 4 \beta_r} T \left[ \frac{1}{2 \mathrm{De}} \left( \frac{M}{p} - 3 \mathrm{B} \frac{p_v}{p} \right) - \mathcal{C}^{\circ}  \frac{1 + \beta_{r}^{\circ} }{1 + \beta_r} J^2 T^{3} \right] \quad ,
\end{equation}
where the reference cooling rate is given by equation \ref{thermal balance}.

\section{Solution to the Induction Equation} \label{induction equation}

In this appendix we derive the the structure of a steady magnetic field in an eccentric disc. The equations for a horizontally invariant laminar flow in a magnetised disc in an eccentric shearing box were derived in \citet{Ogilvie14}. In their coordinate system the induction equation is

\begin{equation}
D B^{\xi} = - B^{\xi} \left( \Delta + \partial_{\varsigma} v_{\varsigma} \right) \quad ,
\end{equation}

\begin{equation}
D B^{\eta} = \Omega_{\lambda} B^{\xi} + \Omega_{\phi} B^{\eta} - B^{\eta} (\Delta + \partial_{\varsigma} v_{\varsigma}) \quad,
\end{equation}

\begin{equation}
D B^{\zeta} = - B^{\zeta} \Delta \quad,
\end{equation}
where $\Omega_{\lambda} = \frac{\partial \Omega}{\partial \lambda}$ and $\Omega_{\phi} = \frac{\partial \Omega}{\partial \phi}$.

In order to rewrite the terms involving derivatives of $\Omega$ we introduce functions $\aleph$ and $\beth$ defined by

\begin{equation}
\frac{\dot{\aleph}}{\aleph} = - \Omega_{\phi} \quad , \frac{\dot{\beth}}{\aleph} = - \Omega_{\lambda} ,
\end{equation}
which in \citet{Ogilvie14} were denoted $\alpha$ and $\beta$. Noting that\footnote{$J$ here being the Jacobian of the $(\Lambda,\lambda)$ coordinate system, as used throughout; as opposed to the Jacobian of the coordinate system of \citet{Ogilvie14} which shares the same symbol. In fact our $J$ is closer to $\mathcal{J}$ of \citet{Ogilvie14}.} $\Delta = \frac{\dot{J}}{J}$ and  $\partial_{\varsigma} v_{\varsigma} = \frac{\dot{H}}{H}$ the $\xi$ and $\zeta$ components of the induction equation become

\begin{equation}
\frac{\dot{B}^{\xi}}{B^{\xi}} + \frac{\dot{J}}{J} + \frac{\dot{H}}{H} = 0 \quad ,
\end{equation}

\begin{equation}
\frac{\dot{B}^{\zeta}}{B^{\zeta}} + \frac{\dot{J}}{J} = 0 \quad , 
\end{equation}
 which have solutions
 
 \begin{equation}
 B^{\xi} = \frac{B^{\xi}_0 (\lambda,\tilde{z})}{J H}, \quad  B^{\zeta} = \frac{B^{\zeta}_0  (\lambda)}{J} \quad ,
 \end{equation}
 where we have additionally made use of the solenoidal condition to show $B^{\zeta}_0$ is independent of $\tilde{z}$. Substituting the solution for $B^{\xi}$ into the $\eta$ component of the induction equation and rearranging we get
 
 \begin{equation}
 \aleph J H \dot{B}^{\eta} +  \aleph J \dot{H} B^{\eta} +  \aleph \dot{J} H B^{\eta} +  \dot{\aleph} J H B^{\eta} + \dot{\beth} B^{\xi}_0 = 0 \quad ,
 \end{equation}
which has the solution

\begin{equation}
B^{\eta} = \frac{\Omega B^{\eta}_0}{n J H} - \frac{\Omega \beth B^{\xi}_0}{n J H} \quad ,
\end{equation}
 where we have used $\aleph \propto \Omega^{-1}$ from \citet{Ogilvie14}. Thus the large scale magnetic field solution in an eccentric shearing box is given by
 
 \begin{equation}
  B^{\xi} = \frac{B^{\xi}_0 (\lambda,\tilde{z})}{J H}, \quad B^{\eta} = \frac{\Omega B^{\eta}_0}{n J H} - \frac{\Omega \beth B^{\xi}_0}{n J H} , \quad  B^{\zeta} = \frac{B^{\zeta}_0  (\lambda)}{J} \quad .
  \label{ideal induction solution}
 \end{equation}
 
 The equation for $\beth$ is given in \citet{Ogilvie14} as
 
 \begin{equation}
 \beth = \frac{3}{2} \left(1 + \frac{2 e \lambda e_{\lambda}}{1 - e^2} \right) \left( \frac{G M}{\lambda^3} \right)^{1/2} t - \frac{\lambda e_{\lambda} (2 + e \cos \theta) \sin \theta}{(1 - e^2) (1 + e \cos \theta)^2} - \frac{\lambda \omega_{\lambda}}{(1 + e \cos \theta)^2} + \mathrm{constant} \quad ,
 \end{equation}
 which in the $(a,E)$ coordinate system is given by
 
 \begin{equation}
 \beth = \frac{3 n t}{2 [1 - e (e + 2 a e_a)] \sqrt{1 - e^2}} - \frac{a e_a}{(1 - e^2)^{3/2}} \frac{2 - e \cos E - e^2}{1 - e (e + 2 a e_a)} \sin E  - \frac{a \varpi_a (1 - e \cos E)^2}{(1 - e^2) [1 - e (e + 2 a e_a)]} + \mathrm{constant} \quad ,
 \end{equation}
 this contains a term that grows linearly in time. This means that $B^{\eta}$ can be expected to grow linearly in the presence of a ``quasiradial'' field.
 
 The ideal induction equation can be written in tensorial form using the operator $\mathcal{D}$ as $\mathcal{D} (B^i B^j) = 0$ \citep{Ogilvie01}. The solution to this equation for a horizontally invariant laminar flow (along with the solenoidal condition) in an eccentric shearing box is given by Equation \ref{ideal induction solution}. In the limit $\tau \rightarrow \infty$ the Maxwell stress obeys $\mathcal{D} M^{i j} = 0$ and the corresponding magnetic field obeys the induction equation. 
 
 The solenoidal condition is not automatically satisfied by the solutions to $\mathcal{D} M^{i j} = 0$ (although it can be imposed). In particular the assumption that the stress has the same vertical dependence as the pressure breaks the solenoidal condition, if $M^{z z} \ne 0$.
 
Finally, when $B^{\xi}_0 = 0$ it is convenient to write the magnetic field in a form which is independent of the horizontal coordinate system used,

\begin{equation}
\mathbf{B} = \frac{B_{h 0} (\tilde{z})}{n J H} \mathbf{v}_{\rm orbital} + \frac{B_{v 0}}{J} \hat{e}_z ,
\label{coord free ish}
\end{equation}
where $B_{v 0}$ is a constant, $B_{h 0} (\tilde{z})$ is a function of $\tilde{z}$ only and $\mathbf{v}_{\rm orbital}$ is the orbital velocity vector.

\section{Derivation of the ideal induction equation model} \label{mag ref state}

We here derive the full set of equations for a horizontally invariant laminar flow in an eccentric disc with a magnetic field. Assume the magnetic field can be split into mean and fluctuating parts:

\begin{equation}
\mathbf{B} = \bar{\mathbf{B}} + \mathbf{b} \quad .
\end{equation}
In order that we have a steady field we require $B^{\xi} = 0$, otherwise there is a source term in the $\eta$ component of the induction equation from the winding up of the ``quasiradial'' ($B^{\xi}$) field. This trivially satisfies the $\xi$ component of the induction equation. We assume the fluctuating field $b$ is caused by the MRI and its effect on the dynamics is captured by the turbulent stress prescription. Thus keeping the mean field only and dropping the overbar the equations for a horizontally invariant laminar flow in a magnetised disc in the eccentric shearing coordinates of \citet{Ogilvie14} are the $\eta$ and $\zeta$ components of the induction equation

\begin{equation}
D B^{\eta} = \Omega_{\lambda} B^{\xi} + \Omega_{\phi} B^{\eta} - B^{\eta} (\Delta + \partial_{\varsigma} v_{\varsigma}) \quad,
\end{equation}

\begin{equation}
D B^{\zeta} = - B^{\zeta} \Delta \quad,
\end{equation}
the momentum equation

\begin{equation}
D v_{\zeta} = - \phi_2 \zeta - \frac{1}{\rho} \partial_{\zeta} \left(p + \frac{B^2}{2 \mu_0}  -  T_{z z} \right) + \textrm{Tension} \quad,
\end{equation}
where $ \textrm{Tension}$ are the magnetic tension terms. The solenoidal condition gives $\partial_{\zeta} B^{\zeta} = 0$, thus $B^{\zeta}$
is independent of $\zeta$ and the magnetic tension terms in the vertical momentum equation are zero. Finally the thermal energy equation is

\begin{equation}
D p = -\Gamma_1 p \left(\Delta + \partial_{\zeta} v_{\zeta} \right) + (\Gamma_3-1) (\mathcal{H} - \partial_{\zeta} F_{\zeta}) \quad ,
\end{equation}
and we must specify the equation of state. Making use of the solutions to the induction equation (Equation \ref{coord free ish}), we obtain an expression for the magnetic pressure,

\begin{equation}
p_M = \frac{B_{h 0}^2 (\tilde{z})}{2 \mu_0 (n J H)^2} v^2 + \frac{B_{v 0}^2}{2 \mu_0 J^2} \quad,
\end{equation}
with  $v^2 = |\mathbf{v}_{\rm orbital}|^2$ is the square of the magnitude of the orbital velocity.
 
The contribution of the vertical component of the magnetic field to the magnetic pressure is independent of the height in the disc (in order to satisfy the solenoidal condition) and makes no contribution to the dynamics of the vertical structure. As such we neglect the vertical component of the magnetic field from this point on. The magnetic pressure simplifies to
 
 \begin{equation}
 p_{M} = \frac{B_{h 0}^2 (\tilde{z})}{2 \mu_0 (n J H)^2} v^2 \quad .
 \end{equation}
On a circular orbit $v^2 = (n a)^2$ so that $p_m \propto (J H)^{-2} \propto \rho^2$ and the magnetic pressure  behaves like perfect gas with $\gamma=2$. Thus, for a magnetised radiation-gas mixture, the magnetic field is the least compressible constituent of the plasma and will be the dominant source of pressure when the plasma is sufficiently compressed. On an eccentric orbit there is an additional source of variability owing to the stretching and compressing of the field by the periodic variation of the velocity tangent to the field lines.

The vertical component of the momentum equation becomes 
 
 \begin{equation}
 \frac{\ddot{H}}{H} = -\phi_2 - \frac{1}{\rho \hat{H}^2 \tilde{z}} \partial_{\tilde{z}} \left(p + \frac{B_{h 0}^2 (\tilde{z})}{2 \mu_0 (n J H)^2} v^2 -  T_{z z} \right) \quad ,
 \end{equation}
where we have used $\hat{H}$ to denote the dimensionful scale height, to distinguish it from the dimensionless scale height $H$. 

We propose separable solutions with
 
\begin{equation}
p = \hat{p} (\tau) \tilde{p} (\tilde{z}) ,\quad T_{z z} = \hat{T}_{z z}  (\tau) \tilde{p} (\tilde{z}) ,\quad \rho = \hat{\rho}  (\tau)  \tilde{\rho}  (\tilde{z}) \quad .
\end{equation}
The dimensionless functions obey the generalised hydrostatic equilibrium which means the pressure obeys
 
\begin{equation}
\frac{d \tilde{p}}{d \tilde{z}} = -\tilde{\rho} \tilde{z} \quad.
\end{equation}

To maintain separability we require the reference plasma beta to be independent of height,

\begin{equation}
\beta_{m}^{\circ}  =  \frac{2 \mu_0 \tilde{p} (\tilde{z}) p^{\circ}}{a^2 B_{h 0}^2 (\tilde{z})} \quad.
\end{equation}
From this we obtain the equation for variation of the scale height around the orbit,

\begin{equation}
\frac{\ddot{H}}{H} = -\phi_2 + \frac{\hat{p}}{\hat{\rho} \hat{H}^2 } \left(1 + \frac{1}{\beta_{m}^{\circ} J^2 H} \frac{P^{\circ}}{\hat{P}} \frac{v^2}{(a n)^2} -  \frac{\hat{T}_{z z}}{\hat{p}} \right) \quad,
\label{ht eq appendix}
\end{equation}
where square of the velocity is
 
 \begin{equation}
 v^2 = (a n)^2 \frac{1 + e \cos E}{1 - e \cos E} \quad .
 \end{equation}
 
The reference circular disc has $f_{\mathcal{H}} = \frac{9}{4} \alpha_s$ as in the hydrodynamic models considered in Paper I. In the reference circular disc, hydrostatic balance is given by
 
 \begin{equation}
 \frac{P^{\circ}}{\Sigma^{\circ} H^{\circ} H^{\circ}} \left(1 + \frac{1}{\beta_{m}^{\circ} } \right) = n^2 \quad.
 \end{equation}
 
 Rescaling Equation \ref{ht eq appendix} by this reference circular disc we obtain
 
  \begin{equation}
 \frac{\ddot{H}}{H} = -(1 - e \cos E)^{-3} + \frac{T}{ H^2} \frac{1 + \beta_r}{1 + \beta_{r}^{\circ} } \frac{\left(1 + \frac{1 + \beta_{r}^{\circ} }{1 + \beta_r} \frac{1}{\beta_{m}^{\circ}  J H T}  \frac{1 + e \cos E}{1 - e \cos E} -  \frac{\hat{T}_{z z}}{p} \right)}{\left(1 + \frac{1}{\beta_{m}^{\circ} } \right)} \quad,
 \end{equation}
 with the rest of the equations proceeding as in the hydrodynamic laminar flow model considered in Paper I. 

\section{Microphysical basis of the nonlinear constitutive model} \label{mag deriv}
 
 Several authors have looked at the possibility of using stochastic calculus as a model of the MRI \citep{Janiuk12,Ross17}. Here we assume the magnetic field satisfies a Langevin equation:
 
 \begin{equation}
 d \mathbf{B} + (\mathbf{B} \cdot \nabla \mathbf{u} - \mathbf{B} \nabla \cdot \mathbf{u} ) d t = - \mathbf{\lambda} d t + \mathcal{F} d \mathbf{X} \quad,
 \end{equation}
 where $\mathbf{X}$ is a Wiener process in the sense of Ito calculus. The left hand side of this equation is the ideal terms in the induction equation, the $- \mathbf{\lambda} d t$ term models damping from resistivity, while the $\mathcal{F} d \mathbf{X}$ represents stochastic forcing by a turbulent electromotive force, where $\mathcal{F}$ is some scale factor controlling the strength of the forcing. In the absence of a mean velocity field $u$, the magnetic field would evolve like a damped Brownian motion.
 
 Introducing $\langle \cdot \rangle$ to denote the expectation value, we have the standard result for the Wiener process $X$,

\begin{equation}
\langle X^i X^j \rangle = g^{i j} t \quad ,
\end{equation} 
where $g^{ij}$ is the inverse metric tensor. So $\mathbf{X}$ is a statistically isotropic vector field. Physically, in this model, turbulent fluctuations act to isotropise the magnetic field. The orbital shear can feed on these fluctuations and induce a highly anisotropic magnetic field that is predominantly aligned/antialigned with the orbital motion. The change in the Maxwell stress can be obtained from Ito's formula,

\begin{equation}
\mu_{0} d M^{i j} = \sum_{n} \frac{\partial M^{i j}}{\partial B^{n}} d B^{n} + \frac{1}{2} \sum_{n m} \frac{\partial^2 M^{i j}}{\partial B^{n} \partial B^{m}} d \langle B^n B^m \rangle \quad,
\label{maxwell change}
\end{equation}
with the partial derivatives given by
 
\begin{equation}
\frac{\partial M^{i j}}{\partial B^{n}} = 2 B^{(i} \delta^{j)}_n, \quad  \frac{\partial^2 M^{i j}}{\partial B^{n} \partial B^{m}} = 2 \delta^{(i}_n \delta^{j)}_m \quad .
\end{equation}
After substituting in these and the equation for $d B$ Equation \ref{maxwell change} becomes
 
\begin{align}
\begin{split}
\mu_{0} d M^{i j} &= - 2 (B^k B^{(i} \nabla_k u^{j)} - B^{(i} B^{j)} \nabla_k u^k ) dt  - 2 B^{(i}\lambda^{j)} d t + 2 \mathcal{F} B^{(i} d X^{j)}  + \sum_{n m} \delta^{(i}_n \delta^{j)}_m  \mathcal{F}^2 d \langle X^n X^m \rangle \\
&= - 2 (B^k B^{(i} \nabla_k u^{j)} - B^{(i} B^{j)} \nabla_k u^k ) dt  - 2 B^{(i}\lambda^{j)} d t + 2 \mathcal{F} B^{(i} d X^{j)} +  \mathcal{F}^2 g^{i j} d t \quad .
\end{split}
\label{stocastic maxwell stress}
\end{align}
Making use of the definition of $\mathcal{D}$, and the fact the Ito integral preserves the martingale property, we can take the expectation of Equation \ref{stocastic maxwell stress} to obtain an equation for the expected Maxwell stress,

\begin{equation}
\mathcal{D} \langle M^{i j} \rangle = - \frac{2}{\mu_0} \langle B^{(i} \lambda^{j)} \rangle +  \frac{1}{\mu_0} \mathcal{F}^2 g^{i j} \quad .
\end{equation}
We should caution that this procedure may not be valid if $\mathcal{F}$ depends on $\mathbf{B}$. Henceforth we shall drop the angle brackets on the modified Maxwell stress and use $M^{i j}$ to denote the expected modified Maxwell stress.

What's left now is to determine appropriate forms for $\lambda^i$ and $\mathcal{F}$. A priori there is no obvious way of directly obtaining these using the underlying physics. However, in a similar vein to \citet{Ogilvie03}, we can place certain constraints on the possible forms of $\lambda^{i}$ and $\mathcal{F}$. In particular on dimensional grounds they both have dimensions of magnetic field over time. $\lambda^{i}$ transforms as a vector and $\mathcal{F}$ transforms as a scalar. Other than this we assume:

\begin{enumerate}
\item Following \citet{Ogilvie03} neither $\lambda^{i}$ or $\mathcal{F}$ directly know about the mean velocity field, although they could know about the various orbital frequencies.
\item $\mathcal{F}$ is non-negative.
\item There is no preferred direction. 
\end{enumerate} 
 
This leaves us to construct a vector and a scalar which have the same dimensions as magnetic field over time from $B^i$, $p_{g}$, $p_{r}$, $\rho$, $\mu_0$ along with the vertical and horizontal epicyclic frequencies and mean motion $\Omega_{z}$, $\kappa$, $n$. Immediately it is apparent that the the only vectorial quantity we have available is the magnetic field $B^{i}$. As such $\lambda^i$ must have the form

\begin{equation}
\lambda^{i} = \frac{B^{i}}{2 \tau} \quad,
\end{equation}
 where $\tau$ is some relaxation time which can depend on the mean field quantities. Next on dimensional grounds $\rho$ cannot appear in either $\mathcal{F}$ or $\tau$ and the other terms must only appear in the combination $|B|$, $\mu_0 p_{g}$ and $\mu_0 p_{r}$ along with the various orbital frequencies. Without loss of generality we can write
 
 \begin{equation}
 \mathcal{F} = \sqrt{\frac{\mu_0 \mathcal{B} p_{\rm v}}{\tau} } \quad,
 \end{equation}
 where $\tau$ is the relaxation time, $\mathcal{B}$ is a dimensionless constant and $p_{\rm v}$ is some reference pressure of the fluctuations. This means our equation for $M^{i j}$ becomes
 
 \begin{equation}
\mathcal{D} M^{i j} = - \frac{1}{\tau} \left( M^{i j} - \mathcal{B} p_{\rm v} g^{i j} \right) \quad,
 \end{equation}
 where we must specify how the relaxation time and fluctuation pressure depend on $M$, $p_{g}$, $p_{r}$ and the orbital frequencies in order to close the model.
 

\bsp	
\label{lastpage}
\end{document}